%% file: main.tex
\documentclass[conference]{IEEEtran}

\usepackage{amsfonts}
\usepackage{graphicx}
\usepackage{xcolor}
\usepackage{soul}
\usepackage[utf8]{inputenc}
\definecolor{vlightgray}{gray}{0.85}
\input{macros}

\newcommand{\din}{D_{in}}
\newcommand{\daug}[1]{\Gamma}

\newenvironment{structure*}{\color{blue}\begin{myenumerate}}{\end{myenumerate}}

\usepackage{marginnote}
\usepackage{balance}
\usepackage{enumitem}

\sethlcolor{yellow}
\newcommand{\update}[3][0em]{#3}

\newcommand{\newcomment}[1]{{{\color{blue} }}}

\usepackage{xspace}
\newcommand{\tinyskip}{\vspace{3pt}}
\newcommand{\mypar}[1]{\tinyskip\noindent\textbf{#1.}\xspace}
\newcommand{\myparwd}[1]{\tinyskip\noindent\textbf{#1}\xspace}

\newenvironment{myitemize}{%
\begin{itemize}[leftmargin=1em, itemsep=.1em, parsep=.1em, topsep=.1em,
    partopsep=.1em]}
{\end{itemize}}

\newenvironment{myenumerate}{%
\begin{enumerate}[leftmargin=1em, itemsep=.1em, parsep=.1em, topsep=.1em,
    partopsep=.1em]}
{\end{enumerate}}

\usepackage{balance}
\usepackage{xcolor} 
\usepackage{tikz}
\usetikzlibrary{fit,calc}
\newcommand*{\tikzmk}[1]{\tikz[remember picture,overlay,] \node (#1) {};\ignorespaces}
\newcommand{\boxit}[1]{\tikz[remember picture,overlay]{\node[yshift=3pt,fill=#1,opacity=.25,fit={(A)($(B)+(.8\linewidth,.8\baselineskip)$)}] {};}\ignorespaces}

\colorlet{pink}{red!40}
\colorlet{blue}{cyan!60}
\colorlet{orange}{orange!60}

\def\HiBl{\leavevmode\rlap{\hbox to \hsize{\color{blue!20}\leaders\hrule height 5\baselineskip depth .5ex\hfill}}}

\def\HiLi{\leavevmode\rlap{\hbox to \hsize{\color{yellow!50}\leaders\hrule height .8\baselineskip depth .5ex\hfill}}}

\begin{document}

\title{\sys: Goal-Oriented Data Discovery}
\author{\IEEEauthorblockN{Sainyam Galhotra, Yue Gong, 
 Raul Castro Fernandez}
 \IEEEauthorblockA{Department of Computer Science,
 The University of Chicago\\
 Email: \{yuegong, sainyam, raulcf\}@uchicago.edu}
 }



\maketitle

\begin{abstract}
Data is a central component of machine learning and causal inference tasks. The availability of large amounts of data from sources such as open data repositories, data lakes and data marketplaces creates an opportunity to augment data and boost those tasks' performance. However, augmentation techniques rely on a user manually discovering and shortlisting useful candidate augmentations. Existing solutions do not leverage the synergy between discovery and augmentation, thus underexploiting data.

In this paper, we introduce \textsc{Metam}, a novel goal-oriented framework that queries the downstream task with a candidate dataset, forming a feedback loop that automatically steers the discovery and augmentation process. To select candidates efficiently, \textsc{Metam} leverages properties of the: i) data, ii) utility function, and iii) solution set size. We show \textsc{Metam}'s theoretical guarantees and demonstrate those empirically on a broad set of tasks. All in all, we demonstrate the promise of goal-oriented data discovery to modern data science applications. 



\end{abstract}

\input{intro}

\input{02prelim}

\input{interventional-queries}

\input{formulation}

\input{04theory}

\input{experiment}

\input{relatedwork}

\input{conclusions}


\bibliographystyle{IEEEtran}
\balance
\bibliography{references}
\end{document}

%% file: macros.tex
\usepackage{amsfonts}
\usepackage[multiple]{footmisc}
\usepackage{booktabs}
\usepackage{dsfont}
\usepackage[utf8]{inputenc}

\usepackage{siunitx}
\usepackage{multirow}
\usepackage{graphicx}
\usepackage{listings}
\usepackage{commath}
\usepackage{epstopdf}
\usepackage{hyperref}
\usepackage{color}
\usepackage{url}
\usepackage{mathtools}
\usepackage{caption}
\usepackage{alltt}
\usepackage{mathrsfs}
\usepackage{float}
\usepackage{subcaption}
\usepackage{graphicx,fancyvrb}
\usepackage[linesnumbered,ruled,vlined]{algorithm2e}
\usepackage{algpseudocode}
\usepackage{amsmath}%
\usepackage{booktabs}

\usepackage{mathtools}
\usepackage{caption}
\usepackage{float}
\usepackage{capt-of}
\usepackage{booktabs}
\usepackage{colortbl}
\usepackage{xcolor}
\usepackage{xfrac}
\usepackage{footnote}

 \usepackage{enumitem}
\usepackage{array}
\usepackage{arydshln}
\usepackage{amsmath,amsfonts}

\definecolor{mygreen}{rgb}{0,0.6,0}
\definecolor{mygray}{rgb}{0.5,0.5,0.5}
\definecolor{mymauve}{rgb}{0.58,0,0.82}

\usepackage{listings}
\lstnewenvironment{VerbatimText}[1][]{
    
    \lstset{fancyvrb=true,basicstyle=\footnotesize,captionpos=b,xleftmargin=2em,#1}
}{}

\usepackage{booktabs}
\usepackage{pgfplots}
\usepackage{pgfplotstable}

\PassOptionsToPackage{hyphens}{url}\usepackage{hyperref}

\usepackage{amsmath}

\usepackage{bbm}

\newcommand{\sys}{\textsc{Metam}}

\newcommand{\ignore}[1]{}

\newcommand*{\rom}[1]{\expandafter\@slowromancap\romannumeral #1@}

\newcommand{\sg}[1]{{{\color{olive} }}}

\newcommand{\option}[1]{{{\color{purple} {}}}}

\usepackage{amsthm}

\newcommand{\RNum}[1]{\uppercase\expandafter{\romannumeral #1\relax}}
\newcommand{\ie}{{\em i.e.}} 
\newcommand{\eg}{{\em e.g.}} 

\newtheorem{defn}{Definition}[section]
\newtheorem{definition}{Definition}

\newtheorem{problem}[defn]{Problem}

\newtheorem{col}[defn]{Colollary}

\newcommand{\proj}[1]{{\Pi}}
\newcommand{\sel}[1]{{\sigma}}

\newcommand{\cut}[1]{}
\newcommand{\eat}[1]{}

\usepackage[multiple]{footmisc}

%% file: intro.tex
\section{Introduction}

Augmenting a dataset by joining it with others can improve the utility of data-driven tasks such as causal inference and supervised machine learning. The abundance of tables in open repositories~\cite{datagov}, lakes in organizations~\cite{halevy2016managing}, and even data markets~\cite{dawex,fernandez2020data} translates into an abundance of \emph{augmentation candidates}. Identifying good augmentation candidates among many is a data discovery problem. To solve this problem, one could use a traditional data discovery system to identify what tables join with the input data, and then, separately, identify what joins increase the task's utility. This \emph{discover-then-augment} approach works when the discovery system returns candidates relevant to the task. Unfortunately, in practice, it is hard to guarantee the discovery system identifies good augmentations because: i) relevant augmentations depend on the task; and ii) analysts may not know what properties make an augmentation relevant, e.g., what features augment the predictive power of a classifier. The disconnection between data discovery and augmentation presents a research opportunity.

We harness that opportunity with a new approach we call \textbf{goal-oriented data discovery}. In goal-oriented data discovery, data discovery is not treated as a one-time process. Instead, it adapts to the task by performing interventions: a process to augment the initial dataset and validate its utility.  \emph{Interventional} queries help to identify augmentations that \emph{cause} an increase in the task's utility, thus steering the discovery process to automatically maximize the task's utility. {This achieves two goals: first, analysts do not need to know what criteria make an augmentation good because the approach is automatic.} Second, any downstream task with a utility function benefits from this discovery approach. The consequences of goal-oriented data discovery are significant; consider the following anecdote. We used goal-oriented data discovery to predict ``housing prices'' in a geographical area. Our technique identified some obvious datasets that a social scientist would have been able to identify using a discovery system, such as ``income of people staying in the neighborhood'' and ``crime stats''. But crucially, it also identified non-obvious datasets correlated with housing prices such as ``presence of grocery stores'' and ``number of taxi trips'' from those areas. Indeed, many sociologists and economists leverage external data to infer causal relationships between attributes of interest~\cite{pope2015walmart,news1,news2}. But they rely on domain knowledge and manual effort to identify those relationships. \emph{Goal-oriented data discovery} paves the way to identify new causal relationships from large data repositories automatically.


A trivial but computationally prohibitive way of solving goal-oriented data discovery is to measure the utility of every augmentation candidate and choose the best. The key technical contributions of our paper focus on developing \emph{interventional querying algorithms} for goal-oriented data discovery that exploit the structure of the data, the utility function, and the solution to adaptively prioritize the candidates for querying.

\mypar{Interventional Queries for Goal-Oriented Data Discovery} Our technical contributions leverage the following properties to optimize the complexity of discovery without loss of quality.

\noindent\ul{Properties of the Data}. A key insight is that \emph{similar} augmentations perform similarly on the downstream task. We exploit this insight by clustering augmentations and judiciously choosing them from different clusters, thus skipping computation. To cluster augmentations, we represent each with a vector of \emph{data profiles}, which are task-independent measures of data and include semantic similarity, correlation, and mutual information, among others. When a combination of data profiles is correlated with the task's utility, clustering narrows down the number of augmentation candidates.

\noindent\ul{Properties of the Utility function}. A task's optimal utility is given by a set of augmentations. Enumerating all subsets is infeasible. Our insight is that when utility functions are monotonic, i.e., utility never decreases with new augmentations, it is possible to find augmentations efficiently, by considering them one by one. And we can make any function monotonic by ignoring augmentations that harm utility using a wrapper around the user-provided task implementation. 

\noindent\ul{Properties of the Solution set}. The optimal set of augmentations is a collection of join paths over different datasets.
Most augmentations are irrelevant for a downstream task, and useful augmentations are a small subset of the candidate set. We leverage this property to prioritize small subsets of join paths over larger ones by using combinatorial testing.

\noindent\ul{\sys{}: Goal-oriented data discovery} We combine the aforementioned properties into an anytime algorithm that finds augmentation candidates by adaptively querying the task. \update{R1A1}{Although goal-oriented data discovery is NP-hard, the algorithm is guaranteed to identify an approximate solution under reasonable assumptions which hold in practice.} We also show the efficiency guarantees of the algorithm, and implement it as part of \sys{}, an end-to-end data discovery approach.

\mypar{Evaluation Results} \sys{} automatically finds useful augmentations among millions of them within minutes. We evaluate \sys{} on large repositories with data from cities in the United States and Kaggle, and for prescriptive analytics tasks such as \emph{what-if} and \emph{how-to} analysis in causal inference, classification and regression in supervised ML, as well as entity linking, and clustering. 

{\noindent \textbf{Outline.}} We present notation and discuss the problem in Section~\ref{sec:background}. We present insights for interventional queries in Section~\ref{sec:interventional}, algorithm in Section~\ref{sec:technical}, and it's analysis in Section~\ref{sec:theory}. We present evaluation in Section~\ref{sec:evaluation}, and related work in Section~\ref{sec:relatedwork}.

%% file: 02prelim.tex
\vspace{-2mm}
\section{Goal-Oriented Data Discovery}
\label{sec:background}

In this section, we introduce notation and problem statement in Section~\ref{subsec:problemdefinition}, and preliminaries in Section~\ref{subsec:challengesandreqs}.


\vspace{-2mm}
\subsection{Problem Definition}
\label{subsec:problemdefinition}

\begin{figure}
\includegraphics[width=0.9\columnwidth]{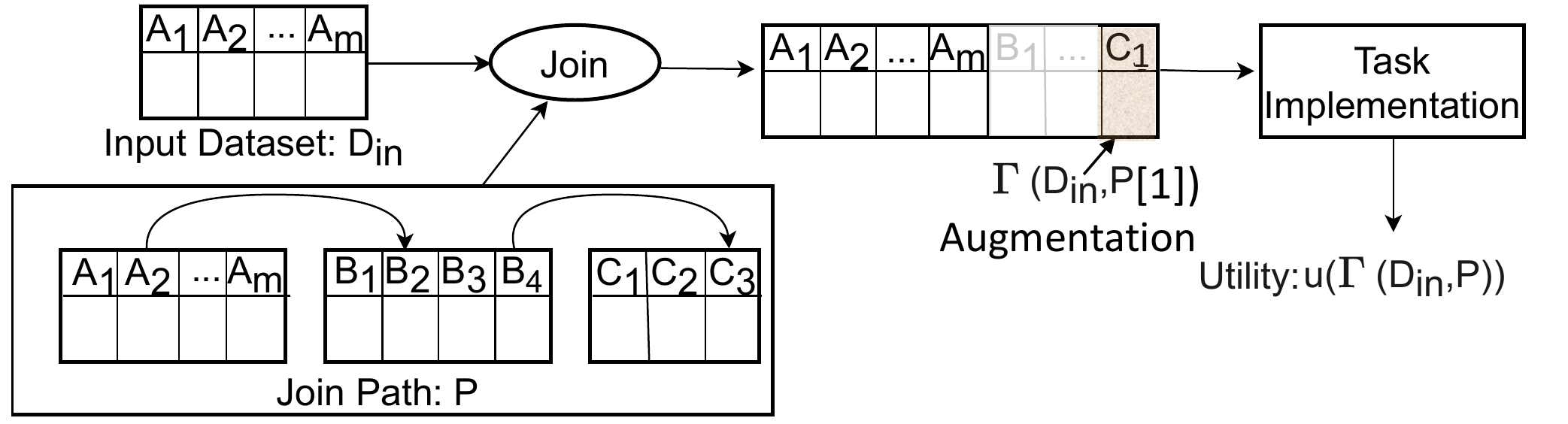}
\vspace{-2mm}
\caption{Notation of a data augmentation pipeline.\label{fig:notation}}
\end{figure}

Figure~\ref{fig:notation} presents a summary of the notation used in the paper. Let $\mathcal{R}(A_1,\ldots,A_m)$ denote a relation schema over $m$ attributes, where $A_i$ denotes the $i^{th}$ attribute. $D$ comprises a schema $\mathcal{R}(A_1,\ldots,A_m)$ and a list of tuples $T$ where each tuple $t\in T$ is a specific instance of the schema. Using this notation, we now define a noisy dataset and a data repository.

\begin{definition}[Noisy Structured Data]
A noisy structured dataset $D$ is characterized by incomplete schema information $\mathcal{R}(A_1,$ $\ldots,A_m)$ where $A_i=\phi$  denotes missing header values and a list of tuples $T$ which may contain less than $m$ values. Additionally, certain  tuples may contain duplicate values.
\end{definition}

\begin{definition}[Data Repository]
A  data repository contains a set of datasets $\mathcal{D}= \{D_1,\ldots, D_n\}$, where different datasets $D_i,D_j$ may contain overlapping values. Additionally, the datasets may lack schema  and contain missing values.
\end{definition}

The different datasets that are joined with the input dataset $\din$ are denoted as a join path, defined below.

\begin{definition}[Join Path]
A join path $P$ is defined as an ordered list of noisy datasets $P\equiv \{D_1,\ldots, D_t\}$ such that datasets $D_j$ and $D_{j+1}$ join for all $j<t$ with each other forming a chain of join operations. The dataset formed by joining these $t$ datasets is considered to be augmented with the original dataset $\din$. 
\end{definition}

The augmentation of a join path $P$ to a dataset $\din$ in denoted by $\Gamma(\din,P)$. Augmenting a join path $P$ with $\din$ can augment multiple new columns, some of which may not be required by the downstream task. To distinguish between these different augmented columns, we define augmentation as the projection of the join consisting of a single column. 

\begin{definition}[Augmentation]
An augmentation $\Gamma (\din, {P[j]})$ is defined as the $j^{th}$ column that is added after joining the path ${P}$ with the input dataset $\din$.
\end{definition}

Note that $\Gamma(\din,P) \equiv \{\Gamma(\din,P[j]), \forall j \in \{1,2,\ldots,l\}\}$, where $l$ is the number of columns after materializing the join-path $P$.
\update{R3O1}{$\Gamma$ considers materializing a join path to add a new attribute. Other types of data augmentation include unions (addition of additional rows), relational embedding based augmentations, spatio-temporal augmentations, etc. Empirically, we observed that join-path based addition of new attributes is commonly used for popular applications. However, our framework extends to any general set of augmentations.  }
A task $t$ is a black box consisting of the analysis algorithm that takes a dataset ($\din$ or an augmented version of $\din$) as input and outputs the performance measure of the task. We call ``query'' to the process of obtaining a task's utility on a dataset. We formally define the utility score of the task as follows.

\begin{definition}[Utility score]
Given a task $t$ that operates on a dataset $D$, the utility score $u_t(D)$ is defined as the objective (or the evaluation metric) of the task when operated with $D$.
\end{definition}

Without loss of generality, we assume that the utility score is normalized $u_t(\cdot)\in [0,1]$ and that higher utility score means better task quality. Different sets of augmentations may improve the task's utility. We seek a \emph{minimal} set that contains only the augmentations that help to maintain high utility.

\begin{definition}[Minimal set of Augmentations]
Given an initial dataset $\din$, a mechanism to compute $u_t(\cdot)$, a set of augmentations $\mathcal{P}=\{P_1[i_1],\ldots,P_j[i_j]\}$ is considered minimal if removing any attribute from  $\Gamma(\din,\mathcal{P})$ reduces task utility.
\end{definition}



\begin{problem}[Goal-oriented discovery]
Given an initial dataset $\din$, a task $t$ that has a mechanism to compute $u(D')$ for any table $D'$, the problem of identifying augmentations that optimize the task is to identify a minimum augmentation set $\mathcal{T}$ such that the augmented dataset has utility $u(\Gamma(\din, \mathcal{T}))\geq \theta.$
\end{problem}

\subsection{\update{}{Task implementation and utility calculation}}
\update{R2O2}{A task takes a dataset as input and returns a utility score that depends, at least in part, on the input dataset. Some examples:}

\begin{myenumerate}
    \item \update{R3 D1.C R3D2}{Predictive analytics: The task trains a machine learning model over an input training dataset. The training code could be simple, e.g., a random forest, or complex, such as an AutoML solution. The utility correspond to a model-relevant metric such as accuracy, F1 score and may also include fairness and robustness metrics.} 
    \item \update{}{Prescriptive analytics involves hypothetical analysis (what-if and how-to queries), causal inference, and explainability as the key components}~\cite{meliou2012tiresias,tableau, powerbi, domo}. \update{}{The utility corresponds to a metric that summarizes the quality of generated explanations, support of identified causal relationship, and total causal effect of attributes.}
    \item \update{}{Data preparation and visualization: These include entity resolution, entity linking, data cleaning and visualization. The utility often corresponds to F-score and accuracy.}
\end{myenumerate}
\update{R3 D1.B}{Note that these are some examples of task implementation and any advances in feature engineering, data transformation and data analysis can be incorporated in the utility function.}

\subsection{Preliminaries and Problem Discussion}
\label{subsec:challengesandreqs}

\update{R3O1}{We use previous data discovery techniques}~\cite{aurum,josie,s4,nargesian2018table} \update{}{to obtain a set of augmentations from large data repositories; we use Aurum}~\cite{aurum}.  \update{}{Although existing solutions may generate noisy candidates (due to the use of approximate techniques and semantic ambiguity), our approach works even when different sources of noise are present.} We also leverage existing profiling techniques~\cite{Santos_2021, lazo, lshensemble,DBLP:journals/corr/abs-2107-04553} to describe the augmentation candidates $\Gamma (\din,{\mathcal{P}})$. For example, a data profile $\langle Corr, C_1,C_2\rangle$ refers to the correlation between columns $C_1$ and $C_2$. We use the term \emph{data profile value} to denote the value of the characteristic/property. A data profile may refer to a single attribute (e.g. domain of an attribute) or multiple attributes (correlation between attributes), \update{}{and these} \update{R2O1}{properties  may be correlated with downstream utility.}

\begin{definition}[Data Profile]
A data profile $X$ is defined as the property of a dataset $D$ such that the tuples in $D$ satisfy $X$.
\end{definition}

\noindent \update{R1O3}{We consider the following data profiles in our implementation.}


\noindent~$\bullet$~\textbf{Correlation and Mutual Information (MI)}: MI is often used to evaluate causal dependence between attributes~\cite{hlavavckova2007causality}. Correlation and MI are also predictors of the quality of attributes in machine learning tasks. This profile estimates the Pearson correlation (MI) of the candidate augmentation $P$ and the attributes of $\din$.  


\noindent~$\bullet$~\textbf{Semantic-embedding based distance} profile captures the semantic similarity between the considered datasets. This profile is computed as the cosine similarity between the embeddings of both datasets constructed from pre-trained models such as BERT~\cite{devlin2018bert}. The dataset embedding is constructed by averaging the embedding vectors of tokens present in the table.

\noindent~$\bullet$~\textbf{Dataset metadata/attributes} profile calculates the similarity between datasets based on their source and attributes. Unlike the semantic embeddings, this profile captures syntactic similarity between attributes and dataset source, which is commonly used to estimate the quality of augmentations~\cite{ver} 

\noindent~$\bullet$~\textbf{Dataset overlap} profile calculates the cardinality of the final dataset identified after augmentation. 



\mypar{Extending to other data profiles} 
\update{R1O3 R1 D1.2}{Metam can be extended with new profiles to cater for new downstream tasks (e.g., anomaly detection}~\cite{brin1997beyond}, \update{}{conditional independence for fairness}~\cite{salimi2019interventional}) \update{}{or to leverage advances in profiling techniques}~\cite{brin1997beyond,chen2018biggorilla, flores2021towards,abedjan2015profiling}.

\subsubsection*{Problem Discussion}

We discuss how various discovery-then-augment baselines fail to solve the problem efficiently:



\noindent~$\bullet$~\textbf{Join Everything}: This technique joins $\din$ with every join path identified. This approach may bring irrelevant attributes that deteriorate the utility. Furthermore, the approach is infeasible in large-scale scenarios with thousands of augmentations.

\noindent~$\bullet$~\textbf{Uniform sampling}: This technique samples join paths out of the identified joinable datasets uniformly at random and uses them to augment $\din$. This approach does not guarantee that the chosen sample will improve the utility score.

\noindent~$\bullet$~\textbf{Join Path overlap ranking}: A common technique is to rank join paths based on the cardinality of the augmented datasets (used by S4~\cite{s4} and Ver~\cite{ver}). This technique identifies datasets that contain fewer missing values, but does not guarantee to optimize the task.
    
\noindent~$\bullet$~\textbf{Using data profiles for selection}: Analysts may use intuition to rank augmentations based on profiles that are expected to be useful. This approach may work if the profiles are accurate and the domain scientist's intuition is correct. But this approach will not work well otherwise, and it relies on accurate estimation of profiles, which are hard to compute in large-scale data repositories, where approximate techniques are often necessary for scalability.


The crux of the problem is that these baselines are unaware of the task, so there is no guarantee that the chosen augmentations will improve utility. Solving goal-oriented data discovery requires an interventional approach that seeks to understand what augmentations \emph{cause} the task's utility to increase.


%% file: interventional-queries.tex
\section{Intervention-based Querying}
\label{sec:interventional}

In this section, we first describe the limitations of baseline techniques and then discuss how the properties of data, utility function, and solution let us design an efficient technique.


\subsection{Baseline Interventional Techniques\label{subsec:baselines}}

Goal-oriented data discovery can be solved by enumerating every subset of candidate augmentations, computing their utility, and choosing the minimal subset that achieves utility of at least $\theta$. With $n$ candidate augmentations, this process may require up to $O(2^n)$ queries (does not finish for $n>30$), so this approach is unfeasible. We consider two improvements:





\noindent~$\bullet$~\textbf{Utility-based selection:} Given a ranking of augmentations (e.g., based on some data profiles) this solution queries, iteratively, in ranking order until the utility obtained is higher than the desired threshold, $\theta$. This solution will be inefficient every time the data profile is not related to the task. And choosing a data profile related to the downstream task is non-obvious for many of the interesting data-driven tasks considered. 

\noindent~$\bullet$~\textbf{Prediction from expert advice:} To avoid selecting a profile a priori, different data profiles can be treated as experts and we can use expert selection techniques to solve the problem. For example, the multiplicative weights update method (MW)~\cite{arora2012multiplicative} estimates the ability of different experts to improve utility. MW guarantees the selection of the best expert in hindsight. In its simplest form, considering each profile as a different expert fails to identify combinations of profiles that best rank augmentations. Enumerating combinations of profiles as experts introduces known combinatorial problems to decision making techniques, including MW~\cite{arora2012multiplicative}.


These two techniques have two shortcomings: i) they do not guarantee finding a minimal set of augmentations that optimize the task's utility; ii) they require $O(n)$ queries, with $n$ indicating the number of candidate augmentations. Furthermore,
no algorithm can find the optimal solution in less than $(2^n-1)$ queries in the worst case (Theorem~2).


\subsection{Towards Efficient Interventional Querying \label{sec:properties}}

{
Even though no technique can be designed to optimize the worst case, we identify several properties of practical scenarios, which help to efficiently identify an optimal (or a close approximation) solution.


}


\mypar{P1 (Optimal solution often contains few augmentations)} Most augmentations are not useful for the task. The number of augmentations, $k$ in the optimal solution is much lower than the total number of augmentations, $k<<n$. Therefore, considering smaller subsets of candidate augmentations has a high chance of identifying the optimal solution. This intuition has previously been studied in the combinatorial testing literature.
We leverage this insight to prioritize the consideration of smaller subsets over large-sized subsets. 

\noindent \underline{What if P1 does not hold?} The scenario where  $k$ is not much smaller than $n$ is not realistic, as augmenting thousands of new attributes blows up the space and worsens overall efficiency. Even in this case, the algorithm identifies the optimal solution as it will eventually consider large-sized subsets.

\noindent \underline{Empirical Validation.} Evaluation over $10$ different tasks considered in Section~\ref{sec:evaluation} identified more than 5000 candidate augmentations ($n>5000$) for each scenario. However, the best solution contained less than $5$ augmentations in $8$ of the cases and less than $25$ in the rest. Therefore, less than $0.5\%$ of the candidate augmentations actually help to improve utility.


\mypar{P2 {(\emph{Similar Datasets are likely to have similar effect on the utility score})}} Two datasets that contain the same set of tuples have the same influence on the utility score. 
Even though the presence of duplicates is an extreme case, we observe that open datasets often contain duplicate information.  We extend this observation to general datasets by considering their similarity. Augmenting two different datasets that have similar profile values are expected to have similar effect on the utility score with more than $0.5$ probability. This property motivates us to consider the dataset properties  to cluster similar augmentations and holistically consider all intra-cluster augmentations for analysis. We use the data profiles introduced in Section~\ref{subsec:challengesandreqs}.
Data profiles are not only useful to cluster the augmentations but also to score them (we refer to these as quality scores). 

\noindent \underline{What if P2 does not hold?} This property is used by \sys{} to prioritize augmentations for querying. In case there is no connection between dataset similarity and their utility score, the identified ordering of augmentations would be the same as a randomized ordering. This would increase the number of required queries to $O(nk)$ in the worst case, where $n$ is the total number of candidate augmentations and $k$ is the number of augmentations in the optimal solution.

\noindent \underline{Empirical Validation.} We compared the difference in utility for augmentations with similarity within $[0.9,1]$ and found that more than 85\% of these have utility difference less than $0.02$. Further, the utility difference increases with reducing similarity.  Therefore, highly similar datasets are generally expected to have similar utility.

\mypar{P3{ (\emph{Monotonicity of the utility function})}} The utility function often satisfies monotonicity with respect to the augmentations, i.e., augmenting new columns to a dataset never worsens the task's utility.  For example, causal inference tasks often estimate the total causal impact of identified attributes, which is monotonic. Accuracy and F-score of a Bayes-optimal classifier~\cite{duda1973pattern} is also monotonic (as Bayes-optimal classifier ignores the newly added feature if it does not help with prediction).
However, certain utility metrics may not be monotonic due to varied reasons e.g. missing values or noise in the newly added attribute. Monotonicity can still be ensured by wrapping the task with a mechanism that ignores an augmentation if it worsens utility. A \textsc{Monotonicity Certification} component (Figure~\ref{fig:metam}) enforces monotonicity in our framework.

\noindent \underline{What if P3 does not hold?} The monotonicity certification component ignores augmentations that worsen utility. This additional check to verify if adding a new augmentation worsens utility may require additional queries. 



\noindent \underline{Empirical Validation.}  We evaluated  monotonicity of the utility for the considered scenarios and identified that causal inference tasks (what-if and how-to) are always monotonic, and  classification tasks are monotonic for more than $60\%$ of the queries. In the remaining $40\%$ of the cases, monotonicity certification component asks additional $\approx 20$ queries and ignores the augmentation that worsens utility to ensures monotonicity.

%% file: formulation.tex
\section{Goal-oriented discovery algorithm}
\label{sec:technical}

\begin{figure*}
    \centering
    \includegraphics[width=0.88\textwidth]{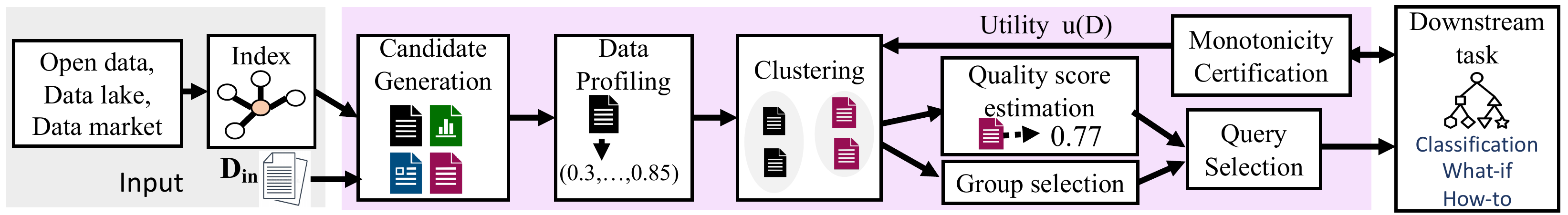}
    \vspace{-2mm}
    \caption{Overview of \sys{}'s architecture.}
    \label{fig:metam}
\end{figure*}

In this section, we give an overview of \sys{} in Section~\ref{sec:algorithm}, and discuss the subroutines in Section~\ref{subsec:subroutines}. Figure~\ref{fig:metam} presents the different components of \sys{}.

\subsection{\sys{} Algorithm Overview}
\label{sec:algorithm}

\sys{} takes as input the initial dataset $\din$, a task $t$ and a collection of datasets $\mathcal{D}$ and outputs a minimal set of augmentations that ensure the task utility of the augmented dataset is at least $\theta$ (see Algorithm~\ref{alg:metam}). The algorithm has two main components: i) \emph{candidate generation and likelihood estimation}; ii) \emph{adaptive querying strategy}. The first component identifies the candidate augmentations and computes the vector of data profiles. The second component alternates between two complementary mechanisms that exploit properties P1--P3 to query the task. Alternating between two procedures allows \sys{} to leverage the best of both techniques and find a solution without assuming anything about the utility function.

\begin{algorithm} 
\scriptsize
\SetKwFunction{AllCombinationOf}{AllCombinationOf} \SetKwInOut{Input}{Input}\SetKwInOut{Output}{Output}
	
	\Input{Dataset $\din$, Data Repository $\mathcal{D}$, utility threshold $\theta$
	} 

	\Output{$\mathcal{T}$: list of augmentations}
	$\mathcal{P}\leftarrow$  \textsc{Generate-Candidates} ($\din,\mathcal{D}$)\label{line1}\\
	\textsc{Evaluate-Profile}($\mathcal{P},\din$)\label{line2}\\
	$\mathcal{T}^*_m,\mathcal{T}^*_c \leftarrow \phi$, $t\leftarrow 1$, $D\leftarrow D_{in}$\\
	$\textsc{JPScore} \leftarrow \textsc{Estimate-Quality-Scores}(\mathcal{P},\din)$\\
	$\mathcal{C} \leftarrow \textsc{Cluster-Partition}(\mathcal{P},\din,\epsilon)$\\
	 \While{ $u(\Gamma(D, \mathcal{T}^*)) < \theta$ \text{and } $u(\Gamma(D, \mathcal{T}_c^*)) < \theta$ }
	 {
	 	$\mathcal{X}, Q\leftarrow \phi$, $i\leftarrow 0$\\
		 \While{ $i < \tau$ \text{ or} $\max_{P\in Q} u'[P] \leq u(D)$\label{line9}}{
		 \tikzmk{A} $P_{max}\leftarrow \max_{P_i\in \mathcal{P}\setminus (\mathcal{X} \cup \mathcal{T}^*)} \textsc{JPScore}(P_i)$\\
		$u'[P_{max}]\leftarrow u(\Gamma(D,\{P_{max}\}))$\\
		
		$\mathcal{X}\leftarrow \mathcal{X}\cup \textsc{Cluster}\{P_{max}\}$\\
		$\textsc{Update-Quality-Scores}(\mathcal{P},P_{max},u')$\\\tikzmk{B}
        \boxit{blue}
		 \tikzmk{A} $\mathcal{P}\leftarrow \textsc{Identify-Group} (\mathcal{C},t)$\\
		   \If{$u(\Gamma(D_{in},\mathcal{P})) > u(\Gamma(D_{in},\mathcal{T}^*_c)) $ }{
		 $\mathcal{T}^*_c\leftarrow \mathcal{P}$, $u'[\mathcal{T}^*_c]\leftarrow u(\Gamma(\din,\mathcal{T}^*_c))$
		}\tikzmk{B}
        \boxit{pink}
		$i\leftarrow i+1, Q\leftarrow Q\cup\{P_{max}\}$\label{line15}
		 }
		 $P_{max}'\leftarrow \arg\max_{P\in Q\cup\{\mathcal{T}^*_c\}} u'[P]$\\
		 \If{$u(\Gamma(D,\{P_{max}'\}))> u (D)$}{
			$\mathcal{T}^*\leftarrow \mathcal{T}^*\cup \{ P_{max}'\}$\\
			 $D \leftarrow \Gamma(D, \{P_{max}'\})$
		}
		 \If{\textsc{Check-Stop-Criterion()}}{
		 break
		 }

	 }
	  $\mathcal{T}^*\leftarrow \arg\max_{T\in \{\mathcal{T}^*,\mathcal{T}^*_c\}} u(\Gamma(\din,T)) $\\
	 $\mathcal{T}\leftarrow \textsc{Identify-Minimal}(\mathcal{T^*},\theta)$\\
	
	\Return $\mathcal{T}$
    \caption{\textsc{\sys{}}}
 	\label{alg:metam} 
\end{algorithm}

\mypar{Candidate Generation and likelihood estimation} First, \sys{} identifies $\mathcal{P}$, the candidate augmentations for $\din$ (line~\ref{line1}) provided by traditional discovery techniques (Section~\ref{subsec:challengesandreqs}). Each augmentation is processed to compute its vector of data profiles (line~\ref{line2}). These profiles are first used to cluster candidate augmentations based on the augmentation similarity; augmentations in the same cluster are expected to impact the utility score similarly (Property P2). This property is used by the querying strategy to choose representative augmentations from each cluster and thereby reduce queries to the utility function. The data profiles are also used to calculate quality score of augmentations to estimate their contributions towards task utility. Intuitively, the quality score is a likelihood estimate that helps to query the relevant augmentations earlier. We explain the scoring technique in the implementation details.
The clusters and quality scores are used to iteratively query the task to evaluate the utility of the augmentation.

\mypar{Adaptive Querying strategy} In each iteration (lines~\ref{line9}--\ref{line15}), the algorithm interleaves \emph{sequential} and \emph{group querying} to evaluate the utility score of the most promising queries. The sequential mechanism (\colorbox{cyan!20}{highlighted in blue}) estimates the quality score of candidate augmentations and chooses the best augmentation by sequentially querying the task.
This procedure processes the augmentations by diversifying the queries across clusters.
Specifically, the quality score of an augmentation is the weighted average of profile scores, where the importance of profiles is used as weights. Profile importance is estimated by evaluating the likelihood that higher profile value indeed achieves higher utility. The quality scores are sorted in a non-increasing order to query the task, with a constraint that at most one augmentation is considered from each cluster. After every query of an augmentation $P$, the identified utility score $u(\Gamma(D,\{P\}))$ is used to update the importance of data profiles and the corresponding scores of other augmentations.  The number of augmentations that are queried before choosing the augmentation with maximum utility is controlled by a parameter $\tau$. 
After $\tau$ queries, the augmentation with maximum utility gain ($P_{max}'$) is added to the solution set $\mathcal{T}^*$ and $D$ is updated to $\Gamma(D,\{P_{max}'\})$ for subsequent iterations. The group mechanism (\colorbox{pink!20}{highlighted in red}) considers subsets of size $t$ (initialized as $t=1$ in line 3) and evaluates its utility. This approach prioritizes useful clusters over less relevant clusters by using  the Thompson sampling~\cite{thompson1933likelihood,russo2018tutorial} mechanism to construct the subsets. We describe this mechanism in Section~\ref{subsec:subroutines}. The value of $t$ is increased when all sets of size less than $t$ have been queried.

\sys{} continues the querying procedure until the utility of augmented dataset is not less than $\theta$ or all augmentations are queried and none of them improve task utility.

\mypar{Minimality check} In the last stage, the best solution among $\mathcal{T}^*$ and $\mathcal{T}_c^*$ is chosen and the identified augmentations are post-processed to identify a minimal set that achieves a utility score of at least $\theta$. This component iteratively removes one augmentation at a time from the solution and evaluates the utility score $u(\Gamma(\din,\mathcal{T}^*\setminus \{T\}))$. If this modified dataset has utility $\theta$ or higher, then $T$ is dropped from the solution set.

\mypar{Stopping Criterion} 
\update{R1 D2.1}{Algorithm}~\ref{alg:metam} \update{}{is an anytime algorithm that stops when the task's utility achieves the threshold $\theta$. If $\theta$ is not provided, the algorithm continues until: i) the search space is explored; or ii) the user finds a good solution.}

 We specify a \textsc{Check-Stop-Criterion} subroutine that tests all user-specified stopping requirements such as time constraint, query budget, solution size constraint, etc.

\subsection{{\textsc{Metam} Subroutines and Generalizations} \label{subsec:subroutines}}

We now delve into \sys{}'s details as used by Algorithm~\ref{alg:metam} to cluster augmentations, estimate quality scores, and propagate the output of a query to other augmentations.

\begin{algorithm}[t]
\scriptsize
\SetKwFunction{AllCombinationOf}{AllCombinationOf} \SetKwInOut{Input}{Input}\SetKwInOut{Output}{Output}
	
	\Input{Augmentations  $\mathcal{P}$,  $\din$, cluster radius $\epsilon$
	} 

	\Output{$\mathcal{C}$: Clusters of Augmentations}
	$c_1\leftarrow \textsc{Choose-Random}(\mathcal{P})$\\
	$S\leftarrow \{c_1\}, \mathcal{C}\leftarrow \{\mathcal{P}\}$\\
	\While{$\textsc{Distance}(\mathcal{C},S)> \epsilon$}{
	$c\leftarrow \textsc{Choose-Farthest}(S)$\\
	$S\leftarrow S \cup\{c\}$, 
	$\mathcal{C}\leftarrow \textsc{Assign}(\mathcal{C},S)$
	}
		\Return $\mathcal{C}$
    \caption{\textsc{Cluster-Partition}}
 	\label{alg:cluster} 
\end{algorithm}

\mypar{\textsc{Cluster-Partition}} We use a distance based clustering technique to ensure that augmentations within the same cluster have all profile values within a distance of $\epsilon$ from their representative.  In other words, Algorithm~\ref{alg:cluster} partitions the space of augmentations into $2\epsilon$ width cubes (of $l$-dimension, where $l$ is the number of considered profiles) where the cluster representatives form the centers. This procedure is often referred to as an $\epsilon$-cover of the augmentations based on the embedding constructed from their profiles. The parameter $\epsilon$ introduces a trade-off between the number of clusters (which impacts the query complexity) and their quality.
Algorithm~\ref{alg:cluster} presents the pseudocode of the clustering algorithm, which  adapts the greedy k-center clustering algorithm~\cite{gonzalez1985clustering} to increase $k$ until all clusters have radius less than $\epsilon$. The distance between augmentations $P_1$ and $P_2$ is calculated as 
{
    $d(P_1,P_2) = \max\limits_{i\in R} d(r_1^i,r_2^i),$
}
where $R$ denotes the set of profiles. 

Given a set of candidate augmentations $\mathcal{P}$ and the initial dataset $\din$ as input, the algorithm chooses cluster representatives (also known as centers) to initialize each cluster and then assigns all augmentations to the respective cluster centers. $S$ denotes the centers and $\mathcal{C}$ denotes a partitioning of $\mathcal{P}$ into the respective clusters. The algorithm initializes the first center by randomly choosing an augmentation and assigning all candidate augmentations to this center (line~1). The subsequent centers are identified by choosing the augmentation that has the maximum distance from its center (lines 3--5). After identifying a new center, all augmentations are re-assigned to the updated set of centers. This farthest identification and reassignment  procedure continue until the farthest augmentation is at most $\epsilon$ away from its center. 

\mypar{\textsc{Quality-Score} Estimation} An augmentation's quality score ranks it according to the expectation of improving task's utility. This score has two components. First, a \emph{profile-based score}, which is a weighted average of their profile values. This score is equivalent to a prior that is estimated from dataset properties. The pipeline is initialized by assigning equal weight to all profiles and these weights are improved with increasing number of queries. Formally, the importance weight of a profile $p$ is defined as the feature importance of $p$ when used to predict the utility of an augmentation.
Second, a \emph{utility-based score}, which indicates the gain in utility score on augmenting a join path. If an augmentation $P$ has been queried previously, the gain in utility is considered as its utility score. If $P$ has not been queried but another join path from the same cluster (say $P'$) has been queried, then, the utility score of $P$ is calculated as the ($1-d(P,P')$) times the score of $P'$, where $d(P,P')$ is the distance between the augmentations.

The quality score of an augmentation is defined as the sum of profile-based score and utility-based score. This score depends on the importance weights of profiles and the utility score of other augmentations in the same cluster. The  \textsc{Update-Quality-Scores} mechanism performs this update.

\mypar{\textsc{Identify-Group}} At a high level, each cluster is characterized by a probability which denotes the likelihood of achieving higher utility on augmentation. Without queries, we do not have an accurate estimate of these probabilities. Querying augmentations from a cluster gives an accurate estimate of the probability but the challenge is to choose between querying a single cluster to get an accurate estimate or diversifying across clusters to explore other options. This follows the traditional explore-exploit dilemma in bernoulli-bandits~\cite{russo2018tutorial}, where the likelihood can be modelled as posterior probability.

We model each cluster as a bandit, where pulling an arm is equivalent to sampling an augmentation. The increase in task utility is the \emph{reward} for a given cluster. We initialize the probability of each arm to $1/|\mathcal{C}|$ and maintain the number of times we get a reward to evaluate the posterior. Each element of the k-sized subset is sampled randomly from the clusters with this probability.

\myparwd{Generalization: What to do when profiles are not useful?} The clustering algorithm identifies groups of augmentations that have similar profile values which impacts Algorithm~\ref{alg:metam}'s query selection. We now analyze the effect of two different types of noises in data profiles on \sys{}. First, if two augmentations that have similar utility do not have similar data profiles, then these augmentations would be placed in separate clusters and would be independently queried to calculate utility. This case is similar to an approach where clustering is not used. It would not affect \sys{}'s quality, but may lead to increased query complexity. Second, when dissimilar datasets have similar data profiles, two augmentations that have similar profiles may not yield similar utility. To handle such settings, we propose a strategy that adapts based on the quality of profiles. Instead of a single query per cluster, \sys{} queries  $\log |C|$ randomly  chosen augmentations from a cluster $C$  to get an accurate estimate of homogeneity with a high probability, \ie{} it checks if the majority of the sampled augmentations  have utility within $(1+\epsilon)$-approximation of the average utility of queried augmentations. If the homogeneity condition does not hold, the quality score estimation ignores  the utility  score component and considers each element as an independent cluster for subsequent iterations of the algorithm.

\noindent \update{R1O3, R2O1}{\textbf{Choosing profiles.}}
\update{}{Developers include any profiles that \emph{may} be correlated with the downstream task. They may not know whether such correlation exists and whether it's strong; they ``cast a wide net'' expecting that some combination of profiles show a correlation. If that is the case, \sys{} will identify it during the search process. How to balance the number and types of profiles included is beyond the scope of this paper;  by default, we include profiles that are effective for machine learning and causal inference tasks. Further, sampling techniques allow to cheapen the computation of profiles.}

\mypar{Impact of $\tau$} $\tau$ determines the number of clusters considered before choosing the augmentation with maximum utility gain. Larger values of this parameter guarantees that the augmentation with maximum utility gain is selected before any augmentation with lower utility gain. Intuitively, this approach identifies a small-sized minimal set of augmentations as compared to \emph{any minimal set}. By default, we choose $\tau = |\mathcal{C}|$  to ensure that at least one  augmentation from each cluster has been queried. Choosing a smaller value of $\tau$ is the same as relying on the quality score to pick the clusters with maximum quality score. In the extreme case, $\tau=1$ is equivalent to querying augmentations in non-increasing order of quality score and selecting any augmentation that improves task utility in the solution set (irrespective of its gain). These settings of $\tau$ should be considered when solution set size need not be optimized. 
We demonstrate the effect of $\tau$ in Section~\ref{sec:quality}.

\mypar{Note}
\update{R1 D1.1}{ The augmentations that do not help to improve utility may be either erroneous (e.g. incorrect join due to incorrect key) or correct but uninformative for the downstream task. \sys{} is robust to handle erroneous augmentations in $\mathcal{P}$.}

\mypar{Risk and Vulnerability}  \sys{} optimizes to choose attributes that maximize utility. Whenever the utility function does not capture the application requirements, e.g., if the outcome is used by a critical decision-making software that cannot be allowed to be fully automated, data scientist can manually verify the augmentations returned by \sys{} and flag the augmentations that do not satisfy their requirements. After flagging these augmentations, \sys{} may have to be rerun to re-optimize for another minimal set of augmentations.

%% file: 04theory.tex
\section{\sys{}'s Performance Guarantees
}\label{sec:theory}

In this section, we show that goal-oriented data discovery is NP-hard (Theorem~1) and no algorithm can identify the optimal solution in less than $2^n-1$ queries. Because the worst case is highly contrived, we analyze \sys{}'s efficiency in practical settings. We show that \sys{} identifies a constant-approximation of the optimal solution in $O(\log n)$ queries. We extend the discussion to noisy settings where the discussed properties (Section~\ref{sec:properties}) may not hold.

\mypar{NP-hardness} A brute force approach to solve goal-oriented data discovery is to consider all subsets of augmentations and choose the one that maximizes utility. To prove the problem is NP-hard, we first consider a decision version of the problem and then show a reduction from the set-cover problem. Consider a decision version that outputs Yes, if there exists a set of size $k$ such that the utility of augmenting the set increases the utility to at least $\theta$.

\mypar{{Theorem 1}}Goal-oriented data discovery is NP-hard.

\noindent \textit{Proof.} Consider an instance of the set-cover problem, where we are given a collection of sets $\mathcal{S}=\{S_1,\ldots, S_m\}$ over a set of elements $U=\{v_1,\ldots,v_n\}$. We want to identify whether there exist $k$ sets in $\mathcal{S}$ such that their union is $U$. 

We show that the set-cover problem is a special case of the decision version of the goal-oriented data-discovery problem. Consider an algorithm that takes a set of augmentations and outputs their utility. For every set $S_i\in \mathcal{S}$, define an augmentation $P_i$. Let the utility function be $u(\mathcal{P}) = |\bigcup_{P_i\in \mathcal{P}} S_i|/n$. If there exists an algorithm that identifies a minimal set of size  $\leq k$ such that $u(\mathcal{P}) = 1$, then there exists a solution of the set-cover problem. This shows that set-cover can be solved optimally, which is a contradiction.\hfill $\qedsymbol{}$

\mypar{Theorem 2}\label{thm:worstcase}
In the worst case, every algorithm identifies an arbitrarily worse solution unless it performs $O(2^n)$ queries.
\noindent \textit{Proof.}
Let $\mathcal{T}$ denote the set of augmentations such that any of the $2^n-1 $ subsets of $\mathcal{T}$ is a valid augmentation. Consider an adversarial utility that outputs  $u(\Gamma(D_{in},T)) = u(D_{in})$ for all queries until the algorithm has queried $2^n -2$ subsets of $\mathcal{T}$. For the last query it outputs  $u(\Gamma(D_{in},T^*)) =\theta$. This utility will require $2^n-1$ queries to identify the optimal solution.\hfill $\qedsymbol{}$

\subsection{Quality Guarantees}

We show that \sys{} guarantees finding the optimal solution even in the worst case when allowed to run for $n^k$ queries. Then, we show that \sys{} identifies a constant-approximation in $O(k)$ queries for the practical scenarios discussed in Section~\ref{sec:properties}.

\noindent \textbf{Optimality Guarantee.} We now show that Metam identifies a set of augmentations that are guaranteed to have a high utility.

\mypar{Theorem 3} \label{thm:guarantee}
If $\exists~ \mathcal{T}^* $ such that $u(\Gamma(\din, \mathcal{T}^*))\geq\theta$, then \sys{}'s output $\mathcal{T}$  satisfies $u(\Gamma(\din, \mathcal{T}))\geq \theta$.
\noindent \textit{Proof.}
The combinatorial testing component of \sys{}'s querying strategy explores all possible subsets of augmentations until a valid solution set is identified. Therefore, the output of \sys{} is guaranteed to achieve utility of at least $\theta$. Since the approach considers all subsets of size atmost $k$ and incrementally increases $k$, it identifies the  solution in $O(n^k)$ queries where $k$ is the size of the optimal solution.\hfill $\qedsymbol{}$

\noindent \textbf{Approximation guarantees.} 
We first show that \sys{} identifies a constant approximation of the optimal solution  within $O(\log n)$ queries. We assume that the optimal solution contains $k$ augmentations, where $k$ is a constant. To prove this result, we first consider the scenario where properties P2 and P3 hold. Later, we relax these assumptions and show that even when similar datasets yield different utility (P2 does not hold) and the utility is non-monotonic (P3 has to be ensured with a wrapper), \sys{} identifies an approximate solution.

\noindent \textbf{Notation.} $\mathcal{C}=\{C_1,\ldots,C_t\}$ denotes the set of cluster centers, $\mathcal{T}^* = \{T_1^*,\ldots, T_k^*\}$ denote the optimal solution, $\mathcal{C}^*$  denote the optimal solution over the cluster centers, and the solution returned by \sys{} is $\mathcal{T} = \{T_1,\ldots, T_k\}$.

\mypar{Theorem 4}
\sys{}'s querying approach identifies a solution with $\frac{1}{\alpha}(1-e^{-\alpha\gamma})-k\epsilon$ approximation in $O(\log n)$ queries.\label{thm:approx}
\smallskip

\noindent \textit{Proof.}
First, we show that the optimal solution over the set of representatives (cluster centers)
is $(1-k\epsilon)$ approximation of the optimal solution over all augmentations $\mathcal{P}$ (Using Lemma~1). Second, we prove that the solution identified by \sys{} is $\frac{1}{\alpha}(1-e^{-\alpha\gamma})$- approximation of the optimal solution over the centers. Combining these two proofs, we get the desired result. Formally,

\vspace{-4mm}
{{
\scriptsize
\begin{eqnarray*}
&&u(\Gamma(\din, \mathcal{T}) \geq {1}/{\alpha}(1-e^{-\alpha\gamma})u(\Gamma(\din, \mathcal{C}^*))\\
&&\geq \frac{1}{\alpha}(1-e^{-\alpha\gamma}) (1-k\epsilon) u(\Gamma(\din, \mathcal{T}^*)\\
&&\geq \left(\frac{1}{\alpha}(1-e^{-\alpha\gamma}) -k\epsilon\right) u(\Gamma(\din, \mathcal{T}^*)
= \left(\frac{1}{\alpha}(1-e^{-\alpha\gamma}) -k\epsilon\right) OPT
\end{eqnarray*}
}}
This shows that \sys{} identifies the approximate solution after $k$ rounds of finding the best augmentation. Additionally, \sys{} tests property P2 (the composition of each cluster) by querying a random sample of $O(\log n)$ queries from each cluster. This step has an added complexity of $O(|\mathcal{C}| \log n)$ to test homogeneity of the clusters. Lemma~2 shows that $|\mathcal{C}| =O(1/\epsilon^l) $, giving an overall complexity of $O(\log n)$.
\hfill $\qedsymbol{}$

\mypar{Lemma 1}\label{lem:representative}
Optimal solution over representatives $\mathcal{C}^*$, $u(\Gamma(\din, \mathcal{C}^*)) \geq (1-k\epsilon) OPT.$

\noindent \textit{Proof.}
To prove this lemma, we apply property of the cluster that the center has utility within a factor of $(1+\epsilon)$ of any other augmentation in the cluster.
\vspace{-4mm}

{
\scriptsize
\begin{eqnarray*}
u(\Gamma(\din, \mathcal{T}_{i}^*)) & = & u(\Gamma(\din, \mathcal{T}_{i-1}^* \cup\{P_{i}\}))
\\
&&\text{Cluster property of adding a new augmentation} \\
&\leq& (1+\epsilon)u(\Gamma(\din, \mathcal{T}_{i-1}^* \cup \{C_{i}^*\})) \\
&\leq & (1+\epsilon)u(\Gamma(\din, (\mathcal{T}_{i-2}^* \cup \{C_{i}^*\}) \cup P_{i-1})) \\
&&\text{Recursively applying the cluster property}\\
&\leq&(1+\epsilon)^k u(\Gamma(\din, \mathcal{C}_{i}^*))
\end{eqnarray*}
}
Therefore, $u(\Gamma(\din, \mathcal{C}_{i}^*)) \geq {(1+\epsilon)^{-k}} OPT \approx (1-k\epsilon)OPT$

\hfill $\qedsymbol{}$

We now bound the number of clusters generated by \textsc{Cluster-Partition} component of \sys{} to show that the number of queries grows linearly in the size of solution set.

\mypar{Lemma 2}\label{lem:cluster}
The number of clusters generated by \sys{} is $O(\frac{1}{\epsilon^l})$, where $l$ is the number of considered profiles.

\noindent \textit{Proof.}
Each join path is represented by $l$ profiles and each profile value is within $[0,1]$. The clustering algorithm chooses a new center whenever the cluster radius is more than $\epsilon$. The l-dimensional space of profiles has unit volume. Therefore, the space can be covered by $O(2^l/\epsilon^l)$ spheres of radius $\epsilon$.
\hfill $\qedsymbol{}$

\mypar{Lemma 3}\label{thm:approx}
$u(\Gamma(\din, \mathcal{T})) \geq \frac{1}{\alpha}(1-e^{-\alpha\gamma})u(\Gamma(\din,\mathcal{T}^*))$, where $\alpha$ and $\gamma$ denote the curvature of $u$ and submodularity ratio, respectively.

\noindent \textit{Proof.}
Here, we analyze a special case where $\gamma = 1$, i.e. the function is always submodular and present the insights of the general result after the proposition.

\vspace{-4mm}
{
\scriptsize
\begin{eqnarray*}
&&OPT=u(\Gamma(\din,\mathcal{T}^*)) \leq u(\Gamma(\din,\mathcal{T}^* \cup \mathcal{T}_i)) \text{ (using monotonicity)}\\
&&\leq u(\Gamma(\din,\mathcal{T}_i)) + \sum_{j=1}^k \left(u(\Gamma(\din,\mathcal{T}_i\cup \{T_j^*\})) - u(\Gamma(\din,\mathcal{T}_i))\right)\\
&&\text{Since $T_{i+1}$ has the maximum marginal gain}\\
&&\leq u(\Gamma(\din,\mathcal{T}_i)) + \sum_{j=1}^k \left(u(\Gamma(\din,\mathcal{T}_i\cup \{T_{i+1}\})) - u(\Gamma(\din,\mathcal{T}_i))\right)\\
&&=u(\Gamma(\din,\mathcal{T}_i)) + k \left(u(\Gamma(\din,\mathcal{T}_i\cup \{T_{i+1}\})) - u(\Gamma(\din,\mathcal{T}_i))\right)
\end{eqnarray*}
}
Therefore, marginal gain of $T_{i+1}$ is atleast $\frac{1}{k} (OPT - u(\Gamma(\din,\mathcal{T}_i))) $.
Let $\delta_i = OPT-u(\Gamma(\din,\mathcal{T}_i))$. Therefore, marginal gain of $T_{i+1} = \delta_i - \delta_{i+1}$.
Using this in the above equation,
$\delta_i-\delta_{i+1}\geq \frac{1}{k}\delta_i$.
Therefore, $\delta_{i+1}\leq (1-1/k)\delta_i$.
Applying this recursively, we get $\delta_{k} \leq (1-1/k)^k\delta_0 \leq \frac{1}{e}\delta_0 = \frac{1}{e}OPT$.
Therefore, $u(\Gamma(\din,\mathcal{T}_k)) = (1-1/e)OPT$.
\hfill $\qedsymbol{}$

This proof is extended from the traditional result for submodular optimization~\cite{nemhauser1978analysis}. In case of general utility functions, the above proof extends using the techniques in \cite{bian2017guarantees}.

\noindent \textbf{Dependence on $\alpha$ and $\gamma$.} \sys{} does not use $\alpha$ and $\gamma$ as input and these notions are used only to characterize the class of utility functions that can be analyzed. The approximation ratio of this result is dependent on the curvature and submodularity ratio of the utility function, which capture the likelihood that the utility function satisfies submodularity. Recent work has studied estimation of these parameters for several real-world functions like linear regression, sparse feature selection, Bayesian A-optimality, determinantal functions, and linear programs with combinatorial constraints~\cite{das2018approximate,bian2017guarantees}. However, it is an open problem to bound the ratios for general functions. 

\noindent \textbf{Empirical Justification.} Experiments in Section~\ref{sec:evaluation} showed that most of the clusters are homogenous, i.e. similar datasets have similar utility. Therefore, \sys{} identifies all useful augmentations within $1000$ queries, which takes more than $10\times$ queries without clustering.

\mypar{Noisy Clusters}
In the case where the identified clusters are not homogenous (property P2 does not hold) and therefore \sys{} ignores all clusters (considers each augmentation as its own cluster) after checking homogeneity in the first iteration. In this case, we show that \sys{}'s quality score estimate is accurate to order the candidate augmentations after $O(\log n)$ queries (proof in the Lemma 4). Therefore, \sys{} identifies useful augmentations within $O(\log n)$ queries. Further, in the worst case when the estimated scores are also inaccurate due to noise in data, \sys{} requires $O(n)$ queries to choose one augmentation, which is repeated $k$ times, yielding an overall complexity of $O(nk)$.

\input{techreport_proof}

%% file: techreport_proof.tex
\noindent \textit{Notation} Let $\beta^* = \langle \beta_1^*,\ldots, \beta_l^*\rangle$ denote the optimal profile scores for quality estimate. Therefore, optimal quality score of an augmentation is $q_i=\sum_{i=1}^l \beta_i^* p_i + \delta$, where $p_i$ is the ith profile value and $\delta_i$ is a Gaussian noise with mean $0$ and variance $\sigma^2$. For this analysis, we leverage the closed form expression of estimated profile scores~\cite{shalev2014understanding}, $\beta = (m\Sigma)^{-1}\sum_i p_iq_i$, where $\Sigma = 1/m \sum p_ip_i^T$. 

\mypar{Lemma 4}
The inferred profile importance, $\beta = \langle \beta_1,\ldots,\beta_l\rangle$ over $m$ augmentation samples has $E[||\beta^*-\beta||^2_2]\leq c\frac{l}{|\mathcal{T}|} $, where $c$ is a constant.

\noindent \textit{Proof.}
We first estimate the difference in $\beta^*$ and estimated $\beta$.
{\scriptsize
\begin{eqnarray*}
q_i &=& \langle \beta^*, p_i \rangle + \delta_i\\
\implies p_i(q_i - \delta) &=&  (p_i p_i^T) \beta^*\\
 \text{Summing over all augmentations}\\
\sum_{i\in \mathcal{T}} p_i(q_i - \delta_i) &=& \sum_i (p_i p_i^T) \beta^*\\
(m\Sigma)^{-1}\sum_i p_i(q_i- \delta_i) &=& \beta^*\\
\implies \beta^* - {\beta} &=& (m\Sigma)^{-1}\sum_i p_i \delta_i 
\end{eqnarray*}
}
Since $||x|| = Trace (xx^T)$
Hence,
{\scriptsize
\begin{eqnarray*}
&&E[||\beta^* - \hat{\beta}  ||_2] = E[\ Trace \left(\left((m\Sigma)^{-1}\sum_i p_i\delta_i \right) \left((m\Sigma)^{-1}\sum_i p_i\delta_i  \right)^T\right)]\\
&&\text{Using }(AB)^T = B^TA^T\\
&&=  E\left(Trace \left(((m\Sigma)^{-1}\sum_i p_i\delta_i ) (\sum_i p_i\delta_i )^T(m\Sigma)^{-T}\right)\right)\\
&&\text{By linearity of expectation}\\
& &= Trace\left( (m\Sigma)^{-1} \left( \sum_{i,j} p_i p_j^T E(\delta_i \delta_j ) \right)(m\Sigma)^{-T}\right)\\
\end{eqnarray*}

\begin{eqnarray*}
&&\text{Since} E[\delta_i\delta_j] = 0, \forall i\neq j\\
& &= Trace\left( (m\Sigma)^{-1} \left( \sum_{i} p_i p_i^T E(\delta_i ^2 ) \right)(m\Sigma)^{-T}\right)\\
&&= Trace\left( (m\Sigma)^{-1} \left( \sum_{i} p_i p_i^T \sigma^2 \right)(m\Sigma)^{-T}\right)
\end{eqnarray*}

\begin{eqnarray*}
&&= Trace\left( (m\Sigma)^{-1} \left( m\Sigma \sigma^2 \right)(m\Sigma)^{-T}\right)= \sigma^2Trace\left( (m\Sigma)^{-T}\right)\\
&&\leq\frac{\sigma^2 Trace(\Sigma^{-1})}{ m}
\leq c\frac{l}{m}
\end{eqnarray*}

}for some constant $c$.\hfill $\qedsymbol{}$

This lemma shows that whenever $\frac{1}{|\mathcal{T}|} = O(1/\log n) = o(1)$, $\beta$ is a close approximation of $\beta^*$.

%% file: experiment.tex
\section{Experimental Evaluation}
\label{sec:evaluation}

In this section, we answer the following research questions.


\noindent~$\bullet$~\textbf{RQ1:} Does \sys{} identify augmentations that improve the utility of the downstream task? (Section~\ref{sec:quality}).

\noindent~$\bullet$~\textbf{RQ2:} Does \sys{} scale with the number of augmentations and data profiles?  (Section~\ref{subsec:scalability}).

\noindent~$\bullet$~\textbf{RQ3:} How sensitive is \sys{} to profile informativeness and parameter choice? (Section~\ref{subsec:ablation})

\mypar{Datasets}  We consider a diverse collection of datasets (schools, taxi, grocery, pharmacy, crime, housing prices, etc.) from two real-world data repositories.

\noindent~$\bullet$~\textbf{Open Data} \update{R1O5}{around 69K datasets comprising datasets from Open Data Portal Watch, which catalogs 262 open data portals such as NYC Open data, finances.worldbank.org, etc.}

\noindent~$\bullet$~\textbf{Kaggle}~\cite{kaggle} contains around $1950$ datasets identified by crawling different competitions.

\begin{table}[t]
\begin{center}
\begin{small}
 
 \begin{tabular}{||c c c c c||} 
 \hline
 Dataset & \#Tables & \#Columns & \#Joinable Columns & Size\\ [0.5ex] 
 \hline\hline
 Open-Data & \update{R1O5}{69K} &  \update{}{29.5M} & \update{}{28.6M} & \update{}{119G}\\ 
 \hline
 Kaggle & 1950 &  91231 & 6.7M &  18G\\ 
 \hline 
\end{tabular}

 \caption{Characteristics of Datasets}
 \vspace{-3mm}
 \label{tab:datasets}
\end{small}
\end{center}
\end{table}

\begin{figure*}
    \centering
    \includegraphics[width=0.95\textwidth]{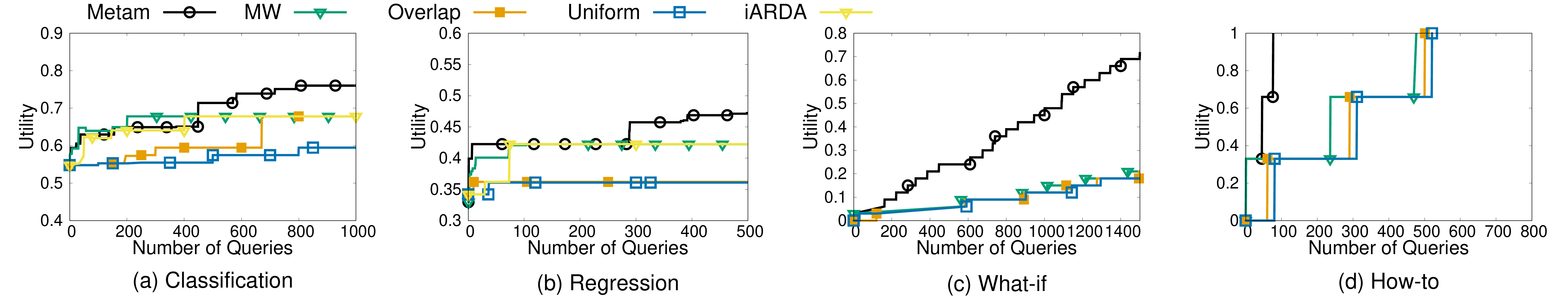}
    \vspace{-2mm}
    \caption{Comparison of \sys{} with baselines (iARDA  is run for supervised ML only) for four tasks on open data repository. }
    \label{fig:regular}
\end{figure*}

\mypar{Settings} We implement \sys{} in  Python and run all experiment on a server with 187GB RAM.  The join path index was computed offline using Aurum~\cite{aurum} with the default parameter settings. Unless specified, $\epsilon$ is set to $0.05$ and $\tau$ is set to the number of identified clusters. We study the effect of these parameters in Section~\ref{subsec:ablation}.
\update{R1O3}{We generate all data profiles (Section}~\ref{subsec:challengesandreqs})\update{}{ on a random sample of $100$ records.}

\subsection{Modern Data Science Toolkit}
\label{sec:quality}

Predictive (machine learning) and prescriptive (causal inference) analytics form the core of modern data analysis. We evaluate \sys{} on both types. 



\noindent$\bullet$~\update{R2O2}{\textbf{Machine Learning}: We consider: i) two classification tasks; ii) a fair-ML task that aims to ensure fairness while maintaining high accuracy; and iii) a regression task. }

\noindent$\bullet$~\update{}{\textbf{Causal Inference}: We consider \emph{what-if} and \emph{how-to} analysis. What-if analysis answers hypothetical reasoning questions, e.g., ``What will be affected if the average revenue increases by 20\%? ''. How-to analysis answers questions such as``How can I achieve revenue more than 20\%?''. These tasks form the core of prescriptive analytics}~\cite{meliou2012tiresias,tableau, powerbi, domo}.


First we measure \sys{}'s performance on these tasks and compare it to other baselines. Second, we demonstrate \sys{} generalizes to other tasks by implementing entity linking, clustering, and fair ML. We demonstrate \sys{} augments the quality of all the above tasks and outperforms baselines without human intervention.


\mypar{Baselines}
No prior work studies goal-oriented data discovery so we adapt prior techniques (see Section~\ref{subsec:baselines}) as baselines. 


\noindent~$\bullet$~Prediction from expert advice (\texttt{MW}): We use the randomized version of the multiplicative weights update algorithm that chooses an expert proportionally to its weight.

\noindent~$\bullet$~Overlap ranking-based approach (\texttt{Overlap}) queries the augmentations in non-increasing order of overlap with $\din$, a mechanism commonly used in prior approaches~\cite{s4}.

\noindent~$\bullet$~Uniform sampling approach (\texttt{Uniform}) samples queries from the candidate set uniformly at random.

\update{R3O3}{We consider ARDA}~\cite{arda}\update{}{ as a task-specific baseline for classification and regression tasks. }


 \subsubsection{Focus on minimizing the size of solution set} We configure \sys{} to find the smallest possible augmentation that boosts the task's utility. Smaller solutions are easier to interpret and thus are highly desirable.
Figure~\ref{fig:regular} shows that \sys{} \emph{outperforms all baselines across all four tasks}. It requires the lowest number of queries yet achieves the highest utility. 


\mypar{Classification (Price)} We predict house prices across different regions in New York, California and Illinois. \update{}{We downloaded around $1000$ records} by searching for randomly chosen zipcodes in these states on Redfin~\cite{redfin}. The task learned a random forest classifier to predict if the house price was low or high. In around 270 queries, the utility achieved by \sys{} rises from $0.69$ to $0.84$. The three most promising augmentations identified by \sys{} include i) presence of Walmart in the same zipcode; ii) number of taxi trips; and iii) crime information (especially for prediction within Chicago). Some of the identified augmentations such as ``presence of a grocery store nearby'' are quite intriguing and we found many studies~\cite{pope2015walmart,caporaso2019taxi} that have shown their importance for predicting house prices. Most importantly, this results were obtained without human intervention beyond pointing \sys{} to the datasets and giving it a task.  



\mypar{Classification (Schools)} The goal is to predict school performance on a standardized test~\cite{arda} \update{R1 D3}{consisting of around 1800 records}.
The task trains a random forest classifier and computes the F-score on a validation dataset as the utility score. The join path index identified more than $9000$ augmentations from the data repository. \update{R1 D1.1}{ We sampled $100$ candidate augmentations and identified that $60\%$ of them are erroneous, e.g. joining datasets with incorrect keys.} In addition to prior baselines, {we adapted ARDA~\cite{arda}, a prior augmentation technique for ML to the interventional setting (denoted by iARDA) where augmentations are queried in decreasing order of ranking returned by \cite{arda}.}

\sys{} outperforms all baselines. It obtains $0.65$ and $0.75$ utility in $200$ and $700$ queries, respectively. This is in contrast to \texttt{MW} and \texttt{iARDA} that require each more than $1200$ queries.
In this dataset, correlation and mutual information of the augmented column with the target attribute are identified as the most informative data profiles by \sys{}. In each iteration, \texttt{MW} considers a single data profile to choose the query, while \sys{} combines all profiles to estimate a quality score. Therefore, \sys{} considers augmentations that have higher mutual information and correlation values early in the discovery process. 


 \begin{figure}
     \centering
     \includegraphics[width=0.9\columnwidth]{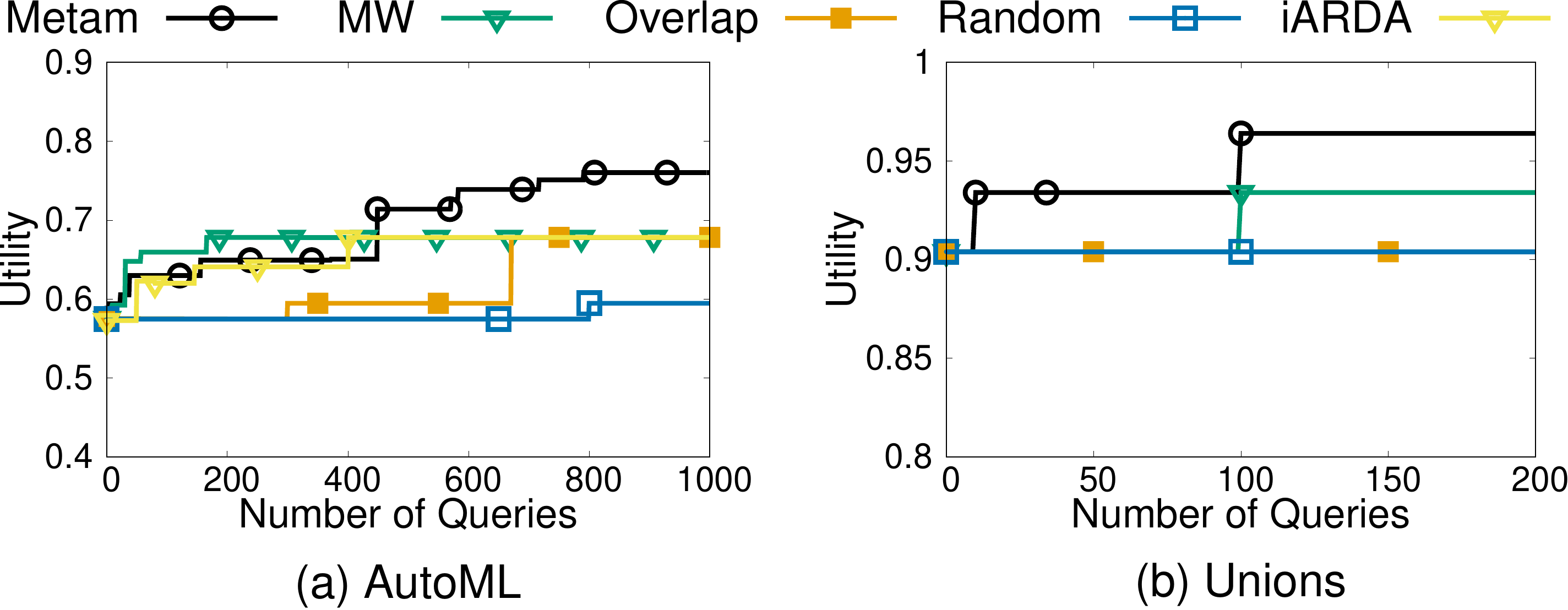}
     \vspace{-2mm}
     \caption{\update{R3O1, R3O2}{Comparison of \sys{} and other baselines for (a) classification with AutoML and (b) Unions (addition of records).} }
     \vspace{-1mm}
     \label{fig:automl}
 \end{figure}

\noindent \underline{\update{R3O2}{Classification with AutoML:}}\update{}{ We considered the schools performance dataset and used three AutoML libraries (Autosklearn}~\cite{autosklearn}, \update{}{TPOT}~\cite{tpot}, \update{}{and PyCaret}~\cite{pycaret}) \update{}{to implement the classifier. Figure}~\ref{fig:automl} \update{}{(a) presents the result for TPOT library. The performance of \sys{} improves the utility of the learned classifier from $0.57$ to $0.75$ in less than $1000$ queries while all other baselines achieve utility of $0.70$. We observe similar results for other AutoML task implementations.}

\noindent \underline{\update{R3O1}{Adding additional records (Unions):}}
\update{}{Prior experiments considered addition of new attributes. In this experiment, we consider a scenario of adding additional records and attributes. Addition of records is often known as unions in data discovery literature. We considered a house rent prediction task for houses in NYC. We used the approach from}~\cite{nargesian2018table} \update{}{to create additional augmentation candidates for unions. Figure}\ref{fig:automl} \update{}{(b) compares the utility of the trained classifier when the base classifier is augmented with additional rows. \sys{} improves utility from $0.90$ to $0.96$ in around $100$ queries while all other baselines reach $0.93$ utility. On further augmenting new attributes to this augmented dataset, utility improved by $1\%$ to $0.974$. } 

\noindent \underline{\update{}{Optimizing the classifier with an expert:}}
\update{R3D3}{We tried to}\update{}{ optimize the classifier by manually performing feature engineering (using trial and error) on the initial dataset for 3hrs. We were able to improve the utility of base classifier from $0.57$ (achieved by AutoML) to $0.62$. However, augmenting external data over this optimized improved its utility to more than $0.75$. This shows that augmenting external data can bring in new attributes that may be highly predictive. Optimizing feature engineering and model learning may not achieve similar gains when the initial set of features are not extremely predictive.}

\mypar{Regression} The goal is to predict number of collisions in NYC using data such as number of daily taxi trips~\cite{arda} \update{R1 D3}{over a dataset containing 350 records.}. The task uses a random forest regressor and computes the mean absolute error (MAE), returning $1-MAE$ as utility.
\sys{} outperforms all baselines. With only $300$ queries, \sys{} reduces MAE from $0.66$ to $0.55$. Other techniques require three times more queries to achieve similar MAE.
Notably, the utility achieved on these tasks is comparable to the values reported in prior work that use these datasets~\cite{arda}, demonstrating effectiveness of \sys{} to achieve competitive quality while generalizing to a wide variety of tasks. We further evaluate predictive analytics (classification and regression) tasks with informative domain-specific data profiles such as feature importance~\cite{arda} and uninformative generic data profiles to understand \sys{}'s flexibility in  Section~\ref{subsec:ablation}.

\myparwd{What-if analysis.} The task takes an input dataset along with a hypothetical update, and outputs the causal impact of the update query on other attributes. We consider an initial table containing \update{R1D3}{SAT scores of $450$ students}~\cite{nycopendata} and ask what attributes would be causally affected if ``critical reading score'' of students is updated. Understanding the attributes sheds light on what affects students' reading score, paving the way for the implementation of interventions. The task implements a causal discovery algorithm~\cite{causallearn} and its quality is measured as the fraction of correctly identified attributes (p-value $\leq 0.05$).


Figure~\ref{fig:regular}(c) shows that \sys{} reaches a utility of $0.70$ in less than $1500$ queries while all other baselines achieve less than $0.25$ utility score. \sys{} reaches utility score of $1$ in around $2000$ queries while \texttt{MW} requires more than $8000$ queries. One of the main reasons for improved performance of \sys{} is the clustering procedure, which allows \sys{} to prune out queries that are expected to have similar utility as that of previously queried augmentations (Property P2). 

\myparwd{How-to analysis}. We want to identify what attributes to update to maximize students' SAT score \update{R1D3}{of 450 students}. The task computes causal dependencies and returns the fraction of correctly identified attributes. The candidate set contains  $240$ augmentations. The optimal solution contains three attributes that have a strong causal impact on the reading score and \sys{} identifies this set of attributes in less than $100$ queries, while all other baselines require more than $500$ queries.

\subsubsection{Allow any size of the solution} We relax the requirement of finding a minimal set of augmentations. The algorithm is allowed to choose the first augmentation that improves task utility rather than choosing the best augmentation. \sys{} consistently outperforms baselines for any stopping criterion. \texttt{MW} requires fewer queries with $\tau=1$ than when optimizing the set of the solution set, but \sys{} still outperforms this baseline. In this relaxed setting, the solution set contains $\approx 9$ augmentations as compared to $2$ augmentations in the experiment that minimizes the solution set size.

\begin{figure}
    \centering
    \includegraphics[width=\columnwidth]{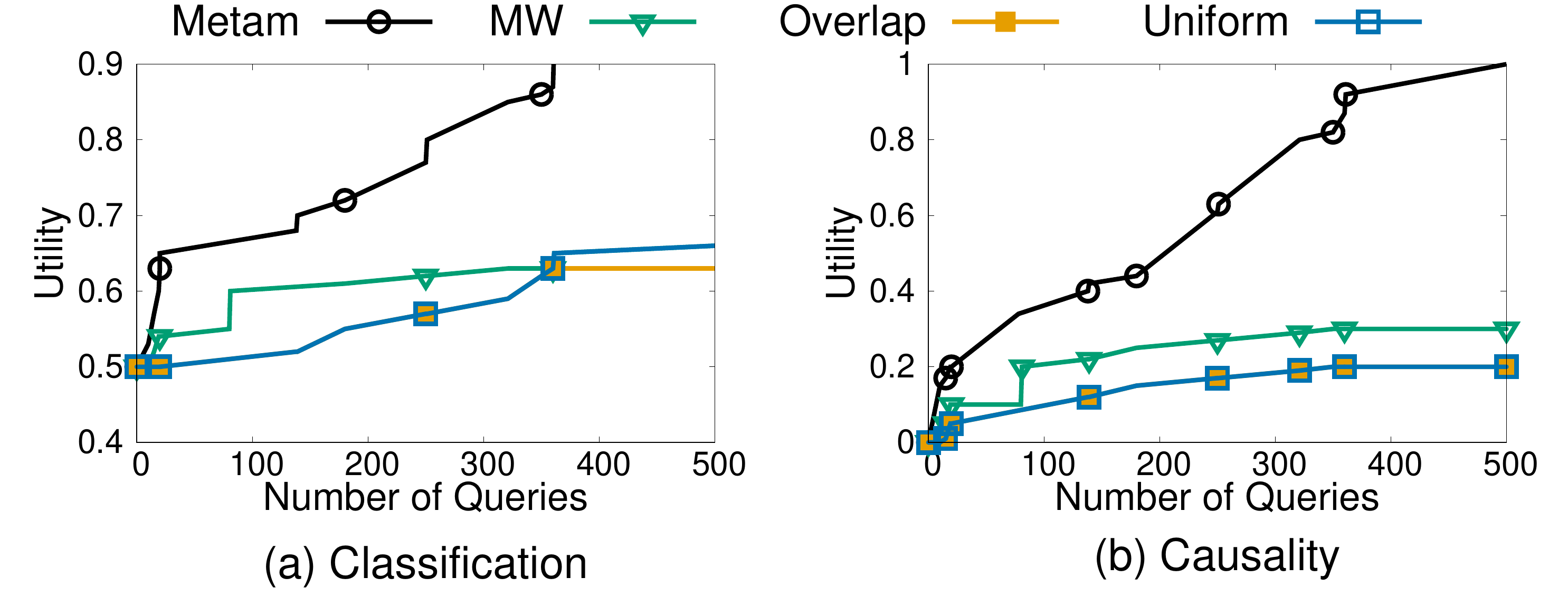}
    \vspace{-7mm}
    \caption{Average result for a semi-synthetic evaluation}
    \vspace{-2mm}
    \label{fig:significance}
\end{figure}

\subsubsection{Results are Significant}

\begin{table}[t]
\begin{center}
\begin{small}
 
 \begin{tabular}{||c | c c c c||} 
 \hline
 Dataset & \sys & MW & Overlap & Uniform\\ [0.5ex] 
 \hline\hline
Schools (C) & \textbf{0.80} & 0.20 & 0.0   & 0.20   \\ 
 \hline
Taxi (C) & \textbf{1} & 0.5  & 0.5  & 0.5 \\ 
\hline
Crime (C) & \textbf{0.90} & 0.20  & 0.1  & 0.1 \\ 
\hline
Housing prices (C)& \textbf{0.75} & 0.25  & 0.0  & 0.25 \\ 
\hline
Pharmacy& \textbf{0.95} & 0.43  & 0.43  & 0.25 \\ 
\hline
Grocery stores  & \textbf{0.92} & 0.37  & 0.10 & 0.17 \\ 
 \hline
\end{tabular}
 \caption{Utility of  \sys{} and other baselines in less than $1000$ queries. Datasets labelled with (C) perform a causal analysis task and others perform data analytics. }
 \label{tab:popular}
\end{small}
\end{center}
\end{table}

In this experiment, we randomly sampled five different augmentations for a randomly chosen dataset in the repository (say $D$) and synthesized a new column in $D$ using these augmentations as i) the prediction attribute for a classification task and ii) the outcome variable for how-to analysis task. This modified version of $D$ was considered as the input table to test the presented techniques. We considered $100$ different instantiations to present average statistics. Figure~\ref{fig:significance} shows that \sys{} consistently outperforms other techniques and is highly effective in quickly discovering useful augmentations. \texttt{MW} is better than other baselines as it is able to identify a useful augmentation whenever it is ranked as one of the highest by any of the data profiles. 

Table~\ref{tab:popular} shows the utility achieved by \sys{} and other baselines for a diverse set of datasets chosen by shortlisting the most accessed datasets on respective data sources. For example, crime data is the most accessed in Chicago's open data, Pharmacy information is one of the most accessed in Pennsylvania's open data portal. Across all datasets, \sys{} is capable of identifying augmentations that achieve the highest utility score. Among baselines, \texttt{MW} vastly underperforms \sys{} even after using considerably more queries.

\subsubsection{Generalization to other tasks}
\label{subsec:generalization}

\myparwd{Entity Linking} aims to link entities~\cite{galhotra2020semantic} in the input dataset to their corresponding entities in knowledge graphs such as Wikidata or DBPedia~\cite{dbpedia}.
We consider a CDC dataset~\cite{kaggle} containing statistics about different cities across the US. Each record contains city name, state abbreviation, \eg{} AL for Alaska. The task is to link all city names to their corresponding entities and evaluate accuracy as the utility metric. 
The task searches for every token on Wikidata and chooses the corresponding entity if it has a unique match. However, city names like Birmingham have multiple Wikidata entities as it is a city in the UK and in the state of Alabama. Augmenting state names to this dataset provides context to the entity linking approach for better linking. We consider the kaggle data repository and identified around $185$ augmentations. \sys{} identified useful augmentations in $4$ queries while \texttt{MW} required $10$ queries and all other baselines required more than $40$ queries.

\mypar{Fair Classification} We consider the credit dataset~\cite{german} to predict income of individuals based on demographic attributes, where age is considered as a sensitive attribute. In addition to the initial dataset, we generate new datasets by transforming some features, following the mechanism employed by prior work~\cite{diazautomated,galhotra2020fair}. The transformed datasets are added to the kaggle repository for discovery. The task implementation internally performs fairness aware feature selection~\cite{galhotra2020fair} to discard attributes that are biased and uses the remaining subset for classifier training. It returns the F-score of the trained classifier as the utility score.  We observe that the attributes which are highly correlated with the target are highly unfair and attributes that are highly fair (have low correlation with the sensitive attribute) do not improve classification accuracy. Therefore, all baselines that rank attributes based on a single profile fail to identify useful datasets within $50$ queries. However, \sys{} considers a weighted average of profile scores, effectively ranks the augmentations and therefore, identifies the minimal set to improve utility in less than $10$ queries.

\mypar{Clustering} This evaluation considers a list of different raw materials and their categories (vegetable, fruit, spices, etc) collected from a health blogging website. The downstream task clusters the products based on their satiety score and returns the additive inverse of the largest cluster radius as the utility. The dataset identifies $8$ augmentations in the repository, one of which augments Optimal nutrient intake (ONI) score of each ingredient. This score is highly correlated with the ground-truth clusters and therefore help to improve clustering quality. Given the small set of candidates, all techniques have a similar performance on this dataset, requiring $\approx 4$ queries.

\vspace{-2mm}
\subsection{Efficiency and Scalability\label{subsec:scalability}}

The number of queries considered in the above experiments are a proxy to evaluate the time taken to run the pipeline. \sys{} is efficient and identifies the set of augmentations in less than $10$ min across all datasets. Most time is spent identifying a candidate set of augmentations and their data profiles. \update{R1 D1.2}{Note that the profiles have varied complexity, e.g. correlation and mutual information are more complex than dataset overlap. Out of $10$ minutes, roughly $4$ minutes are spent on generating data profiles for the candidate augmentations. } For a fixed number of queries, \texttt{MW} requires the same amount of time than \sys{}. The other baselines run slightly faster, taking $7$ minutes.

\begin{figure}
     \centering
     \includegraphics[width=0.9\columnwidth]{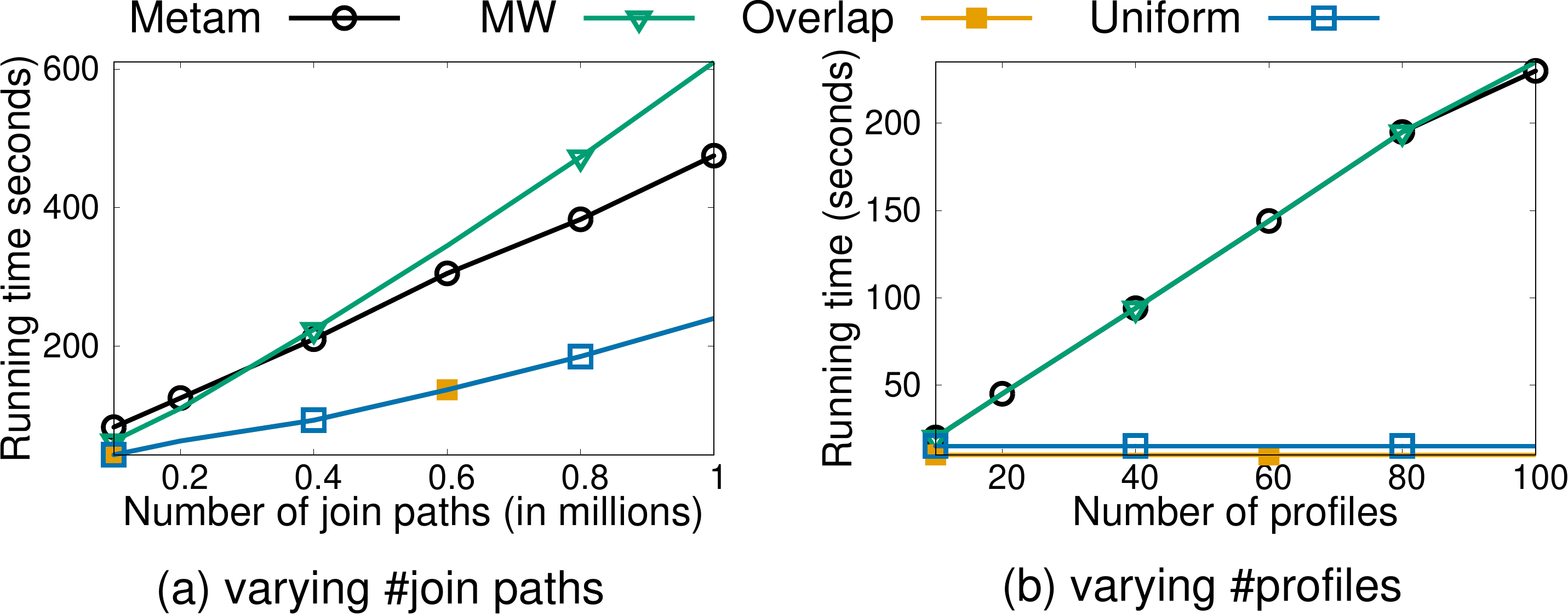}
     \vspace{-2mm}
     \caption{Execution of \sys{} and other baselines for (a) varying the set of joinable datasets and (b) data profiles. }
     \vspace{-4mm}
     \label{fig:scalability}
 \end{figure}

To evaluate scalability, we vary the number of candidate augmentations and data profiles and measure the time needed to run $1000$ queries. 
Figure~\ref{fig:scalability} 
(a) shows that all techniques scale linearly with the number of joinable datasets in the repository. The time taken by \texttt{MW} increases faster than \sys{} due to $O(n\log n)$ time taken to sort augmentations. In contrast, \sys{} clusters augmentations in $O(n)$ time and uses identified clusters for efficient computation. Although \texttt{Overlap} and \texttt{Uniform} are faster, they vastly underperform the other baselines, as previously shown. Figure~\ref{fig:scalability} (b) shows that \sys{} and \texttt{MW} scale linearly with the number of profiles, while the time taken by \texttt{Overlap} and \texttt{Uniform} does not change as they do not use any of the input profiles. In summary, \sys{} processes $1M$ augmentations in less than 10 minutes and scales linearly with the number of augmentations.
 
\begin{figure}
    \centering
    \includegraphics[width=0.95\columnwidth]{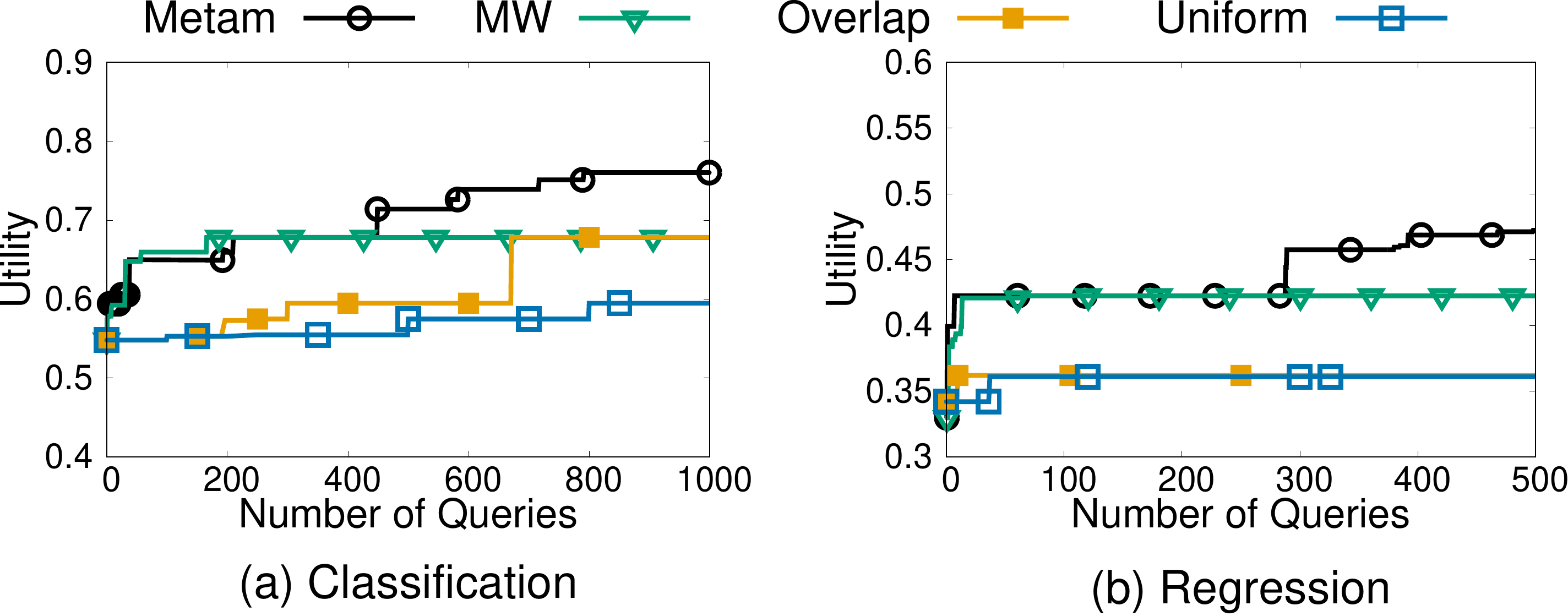}
    \vspace{-2mm}
    \caption{Comparison of \sys{} with other techniques  using task-specific profiles from \cite{arda}.}
    \label{fig:arda}
\end{figure}

\subsection{\update{}{Effect of Profile Informativeness and Parameters}}
\label{subsec:ablation}

\update{}{In this section, we first evaluate the impact of incorrect  candidate augmentations (false positives) on overall performance of \sys{}. We then evaluate the impact of (a) adding informative profiles (b) uninformative ones, and (c) removing informative profiles on \sys{}'s ability to improve downstream utility and overall efficiency. We then evaluate an extreme scenario where all profiles are uninformative. Lastly, we perform an ablation analysis to understand the impact of different components of \sys{} and their parameters.}

\begin{figure}
    \centering
    \vspace{-1mm}
    \includegraphics[width=0.9\columnwidth]{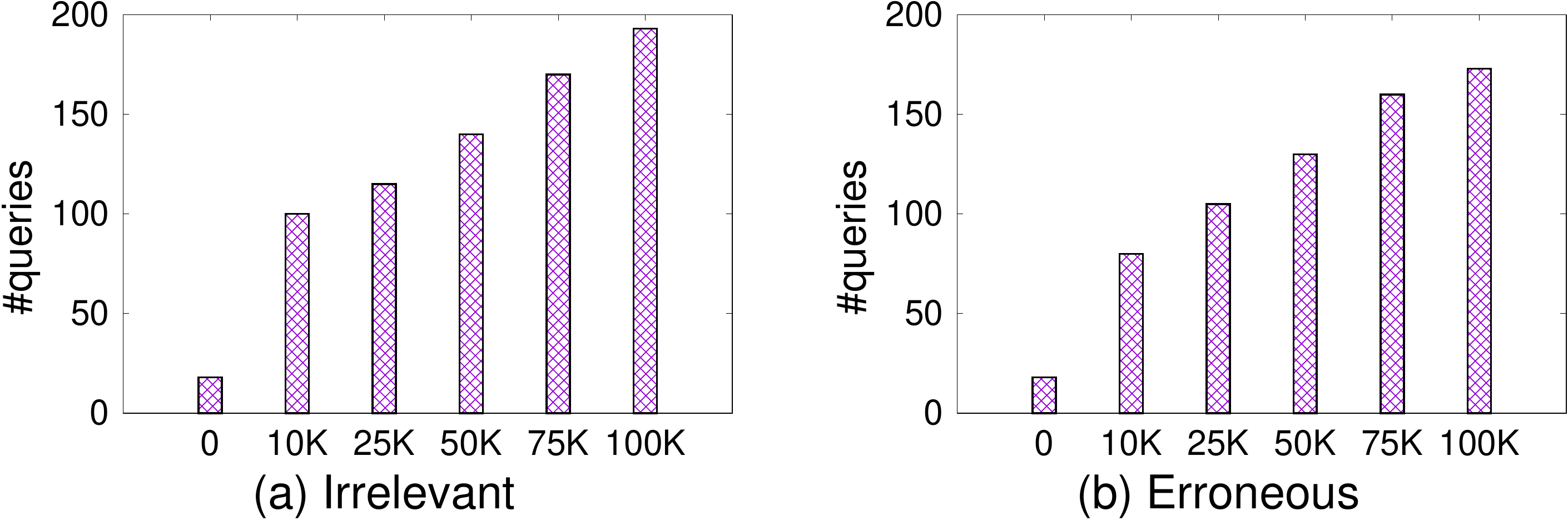}
    \vspace{-2mm}
    \caption{\update{R1 D1.1}{Number of queries required to identify ground truth  with varying irrelevant and erroneous augmentations.}}
    \label{fig:erroneous}
\end{figure}

\noindent \textbf{\update{R1 D1.1}{Do incorrect augmentations impact \sys{}'s performance?} }\update{}{ In this experiment, we considered a single ground-truth augmentation for classification and regression datasets. Figure}~\ref{fig:erroneous} \update{}{(a) fixes erroneous augmentations (incorrect join path identified by candidate generate algorithm) to $100$ and varies correct but irrelevant augmentations (do not help task utility) and Figure}~\ref{fig:erroneous}\update{}{ (b) fixes the irrelevant augmentations but varies erroneous augmentations. We observed that \sys{} identified the ground truth augmentation and discards all incorrect and irrelevant augmentations within $200$ queries. The number of required queries increases with increasing number of irrelevant or erroneous augmentations. }

\begin{figure}
\vspace{-5mm}
    \centering
    \includegraphics[width=0.9\columnwidth]{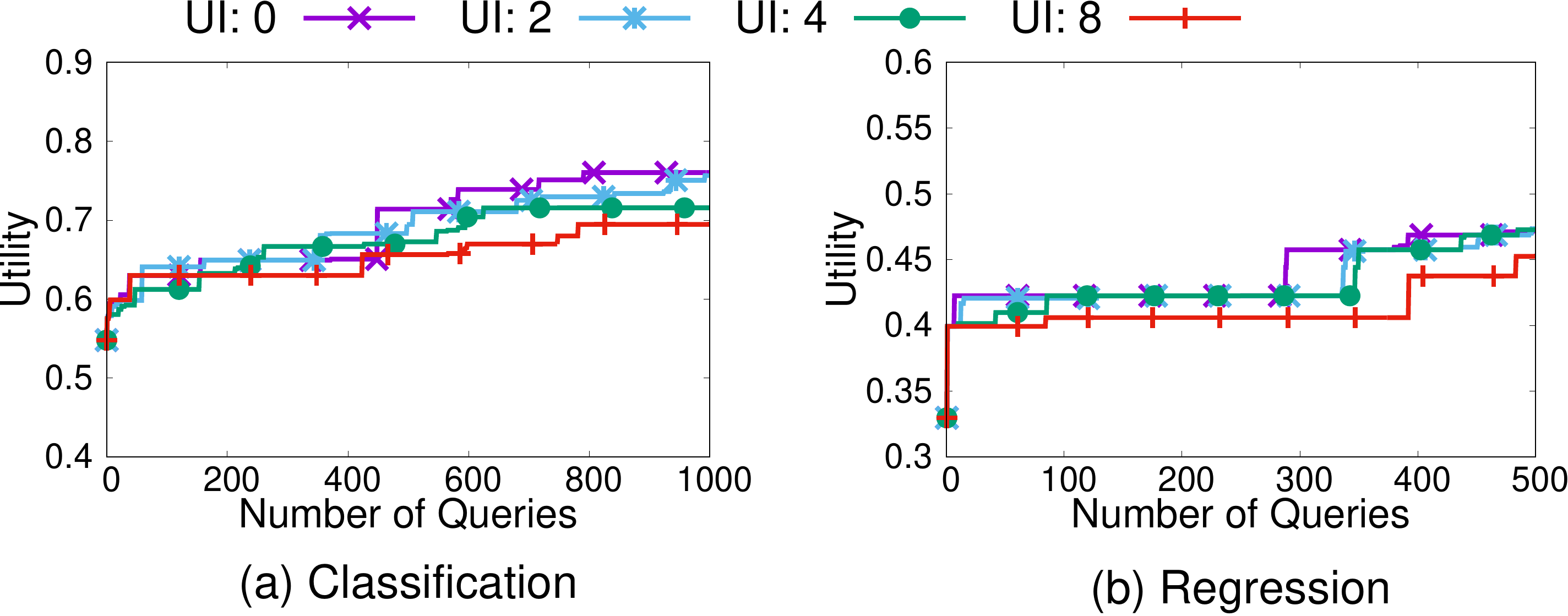}
    \vspace{-3mm}
    \caption{\sys{} performance with varying number of uninformative profiles. The number of informative profiles is fixed to be $5$ and uninformative are denoted by UI.}
    \label{fig:addnoise}
\end{figure}

\mypar{\update{R1 D1.2}{Adding informative profiles}} \update{}{In this experiment, we add informative task-specific data profiles 
using Arda}~\cite{arda}. Figure~\ref{fig:arda} 
\update{R2O1}{shows that \sys{} requires fewer queries than \texttt{MW} and other baselines. In fact, \sys{} achieves utility of $0.68$ in less than $200$ queries, as opposed to more than $400$ queries when these specialized data profiles are not used. Informative profiles boost \sys{}'s performance.} 


\mypar{\update{R1 D1.2}{Adding uninformative profiles}}\update{}{This experiment considers the initial set of profiles (as described in Section}~\ref{sec:technical}) \update{}{and generates additional uninformative profiles.
Figure}~\ref{fig:addnoise} \update{}{shows that \sys{} achieves similar utility gains across all settings and adding noisy profiles does not affect the quality of the solution because \sys{} learns to ignore them at the cost of running a few more queries.}

\begin{figure}
    \centering
    \includegraphics[width=0.9\columnwidth]{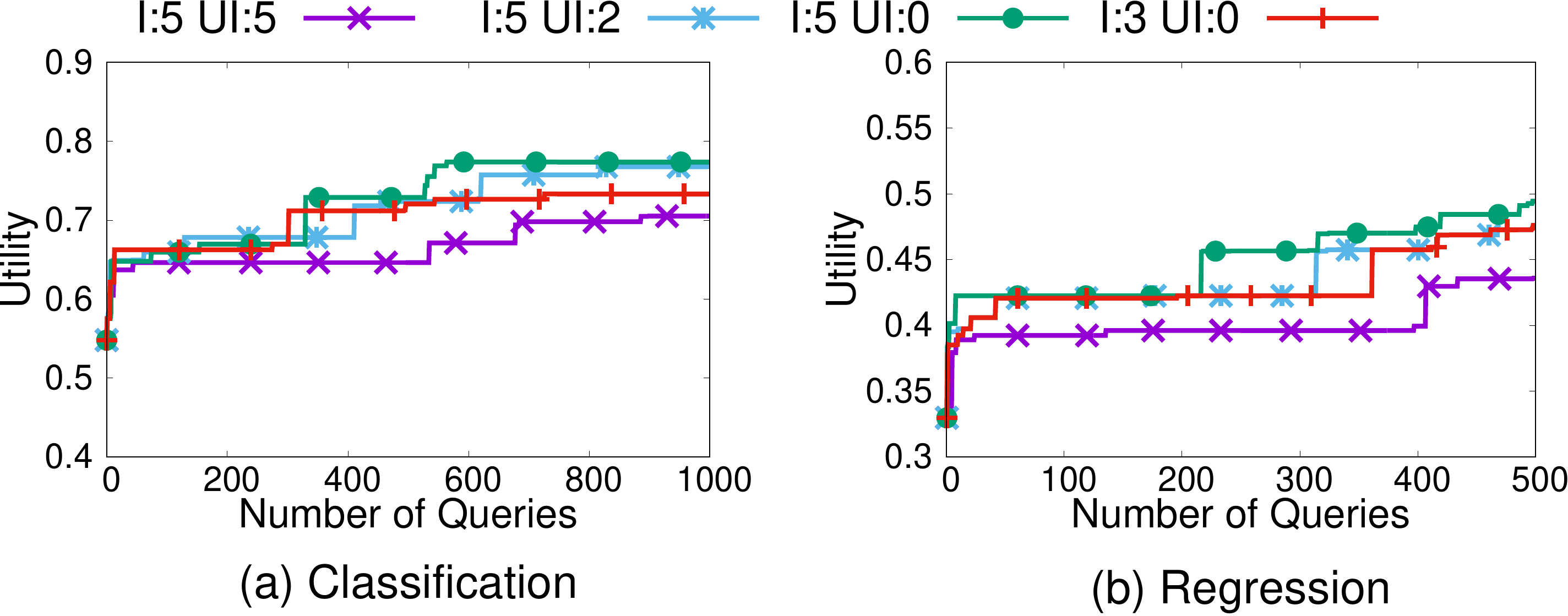}
    \vspace{-2mm}
    \caption{Effect of removing profiles on \sys{} quality vs query tradeoff. I denotes the number of informative profiles.}
    \label{fig:signalremoval}
\end{figure}

\mypar{\update{}{Removing  profiles}}\update{}{ We consider the classification and regression tasks with $10$ data profiles, $5$ of which are uninformative. First, the uninformative profiles are removed, followed by removal of other informative profiles. Figure}~\ref{fig:signalremoval} \update{}{shows that removing irrelevant features (until the number of removed features is less than $5$) improves the progressive growth in task utility with respect to the number of queries. On further removing any of the relevant profiles, the number of required queries to reach a similar utility increases.} 

\noindent \textbf{\update{R2O1}{What if all profiles are uninformative?}} \update{}{ In this experiment, we considered the schools dataset for classification where all profiles are generated randomly.}\update{R1 D1.2}{ In this case, the quality score estimated by \sys{} is similar to a random ordering. However, \sys{} identifies the optimal set of augmentations, but the number of queries is similar to that of uniform baseline. } 

\begin{figure}[ht]
    \centering
    \includegraphics[width=0.9\columnwidth]{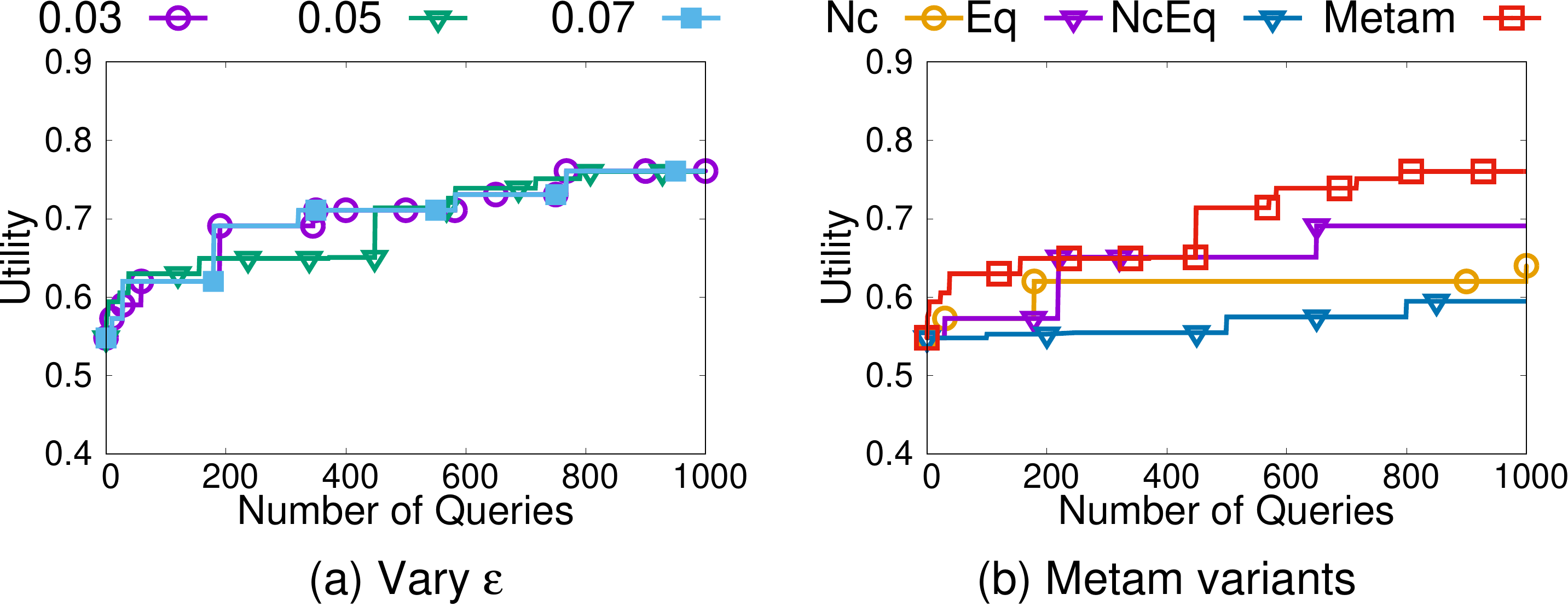}
    \vspace{-2mm}
    \caption{Ablation analysis $\epsilon$, and \sys{} variants.}
    \label{fig:ablation}
\end{figure}

\mypar{\sys{} and its variants} 
Figure~\ref{fig:ablation} (b) compares \sys{} with three variants, \update{R1 D1.3}{(i) Eq: This variant is equivalent to ignoring Thompson sampling procedure and ranks each cluster with equal importance.  (ii) Nc: considers each augmentation as a different cluster, which is equivalent to ignoring property P2. (iii) NcEq: considers each augmentation equally important and ignores clustering. \sys{} outperforms all variants because Eq and NcEq process the candidate set of augmentations in a random order while Nc wastes queries on redundant augmentations that are not expected to help with the downstream utility (which are clustered together by \sys{}).} Therefore, considering clustering and ranking mechanisms together allows \sys{} to prioritize important augmentations and efficiently ignore redundant ones.

\vspace{-1mm}
\mypar{Impact of $\epsilon$} Figure~\ref{fig:ablation} (a) measures the quality of \sys{} when varying $\epsilon$ for clustering. 
Increasing the value of $\epsilon$ reduces the number of clusters in the dataset and increases the distance between intra-cluster augmentations. Therefore, the property P2 may be violated in such a case and \sys{} may ignore the cluster structure for quality score estimation. On the contrary, choosing a smaller value of $\epsilon$ means that all augmentations that are present in the same cluster are very similar with respect to the considered augmentations and therefore, all augmentations within a cluster may have similar effect on the utility of the downstream task. Figure~\ref{fig:ablation} shows that \sys{}  the number of queries do not change drastically on varying $\epsilon$.

%% file: relatedwork.tex
\section{Related Work}
\label{sec:relatedwork}



\mypar{Data Discovery Approaches} Data discovery systems and libraries \cite{ver, infogather, aurum, auctus, lazo, lshensemble, Santos_2021, DBLP:conf/edbt/EsmailoghliQA21,nargesian2018data,nargesian2020organizing} compute join paths from data repositories automatically. Hence, given a query dataset, these systems tell users what other datasets join with the query dataset. Many modern techniques allow users to interactively search useful joinable datasets~\cite{zhang2020finding} by specifying requirements in the form of data profiles or summaries and explanations~\cite{santos2019visus}.  However, without the information of user's goal, it is difficult to identify which join path is useful to the downstream task, as we have demonstrated in the evaluation section.



\mypar{Human-in-the-Loop Data Integration}
Human-in-the-loop approaches \cite{DBLP:journals/pvldb/Li17,  DBLP:conf/cikm/ZhuangLZF17, DBLP:journals/tods/BonifatiCS16} utilize user feedback to steer the search process. These techniques have been applied to problems ranging entity resolution~\cite{vesdapunt2014crowdsourcing,firmani2016online,wang2013leveraging,gokhale2014corleone}, join selection, and entity alignment, among others~\cite{li2017human}. These approaches' focus is to reduce the effort necessary to find solutions. In contrast, \sys{} focuses on automatic data discovery and augmentation that does not require user intervention.


\mypar{Data and Machine Learning (ML)} 
\update{R3D4}{There are data augmentation techniques} \cite{arda, DBLP:journals/pvldb/00010MT21,kumar2016join,liu2021automatic,chai2022selective,zhao2022leva} \update{}{designed for ML tasks that either select attributes and data points that are useful for prediction or learn relational embeddings to embed data into a vector space.} These techniques are not interventional, and inherently rank available datasets based on their importance, which can be used as data profiles for \sys{}. Therefore, the contributions of this work are orthogonal to those of prior techniques that rank features or augmented datasets based on their importance.  
Additionally, \textsc{Metam} is not restricted to machine learning tasks. Instead, it implements \emph{goal-oriented data discovery} and consequently can steer data discovery for any task for which developers can provide a function that computes the utility score. Furthermore, \textsc{Metam} dynamically changes the weights of different profiles, making it more robust to data quality issues in join paths and profiles.



\mypar{Query-Driven Tasks} Query-Driven approaches \cite{DBLP:journals/pvldb/AltwaijryKM13, DBLP:journals/pvldb/AltowimKM14} focus on reducing the calculation of tasks like entity resolution based on the requirements of a concrete query. These approaches solve the efficiency problem in an online setting and require a well-defined query a priori. However, it is impossible to have a well-defined query in data discovery problems, i.e. usually analysts do not know clearly which portion of the data collection is useful for their task.


%% file: conclusions.tex
\section{Conclusions}
\label{sec:conclusions}

We proposed a novel approach to perform goal-oriented data discovery that runs interventional queries to efficiently identify useful augmentations. Our approach leveraged properties of the data, utility function and the solution set to optimize the querying strategy. We analyzed the optimality and the query complexity of our approach.
Empirical analysis on a diverse set of datasets from different application scenarios and comparison with baselines demonstrate versatility and effectiveness to efficiently select useful augmentations.
